\documentclass[12pt]{article}

\usepackage{graphicx}
\makeatletter
 
  \@addtoreset{equation}{section}
 \makeatother
 
\newcommand{\nn}{\nonumber}
\newcommand{\beq}{\begin{eqnarray}}
\newcommand{\eeq}{\end{eqnarray}}

\newcommand{\mbf}[1]{\mbox{\boldmath$#1$}}
\newcommand\el{{\rm e}}
\newcommand{\nrmk}{|\mbox{\boldmath $k$}|}

\def\bk{{\mbox{\boldmath$k$}}}

\def\bp{{\mbox{\boldmath$p$}}}

\makeatletter
 
  \@addtoreset{equation}{section}
 \makeatother

 \renewcommand{\appendix}{%
\renewcommand{\section}{%
 \secdef\Appendix\sAppendix}%
 \setcounter{section}{0}%
\renewcommand{\thesection}{\Alph{section}}%
}

\newcommand{\Appendix}[2][?]{
\refstepcounter{section}%
\addcontentsline{toc}{appendix}%
{\protect\numberline{\appendixname~\thesection} #1}%
{\flushleft\Large\bfseries\appendixname\ \thesection\ \ 
#2\par
}%
\sectionmark{#1}\vspace{\baselineskip}}

\newcommand{\sAppendix}[1]{
{\flushright\large\bfseries\appendixname\par
}%
\vspace{\baselineskip}} 
\begin{document}

\begin{center}
{\large  \bf Effect of negative energy components for two-nucleon systems in the relativistic framework
with the separable ansatz}
 
 \vspace*{1cm}
Y. Manabe, A. Hosaka and H. Toki\\
\vspace*{2mm}
 {\it Research Center for Nuclear Physics (RCNP), Osaka University\\
 Ibaraki, Osaka 567-0047, Japan}\\
 %\today
\end{center}
\vspace*{0.2cm}

\abstract{
We investigate the electromagnetic properties
of the deuteron such as the charge and magnetic form factors
by solving the Bethe-Salpeter equation (BSE) with the separable ansatz.
In solving the deuteron bound state solution to the BSE, 
we include negative energy components of $P$-wave in addition to the
$^3S_1$ and $^3D_1$ states of positive energy, employing a rank IV separable ansatz. 
We found that the inclusion of the negative energy components improves systematically
the electromagnetic properties which are not described 
in the conventional non-relativistic impulse approximation.    
}

%\vspace*{1cm}
 
%======================================================
\section{Introduction}
\begin{figure}[h]
\centering
\begin{minipage}{15cm}
\centering
\includegraphics[width=7cm]{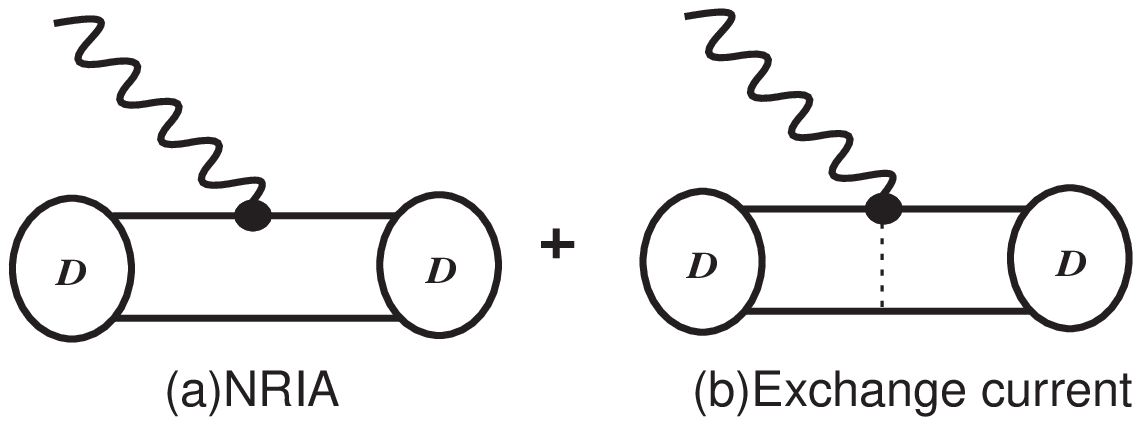}
%\vspace{-0.5cm}
\caption{Feynman diagrams for non-relativistic impulse approximation (NRIA).}
	 \label{fig:nrd}
\end{minipage}
\\
\vspace{1.0cm}
\begin{minipage}{15cm}
\centering
\includegraphics[width=13cm]{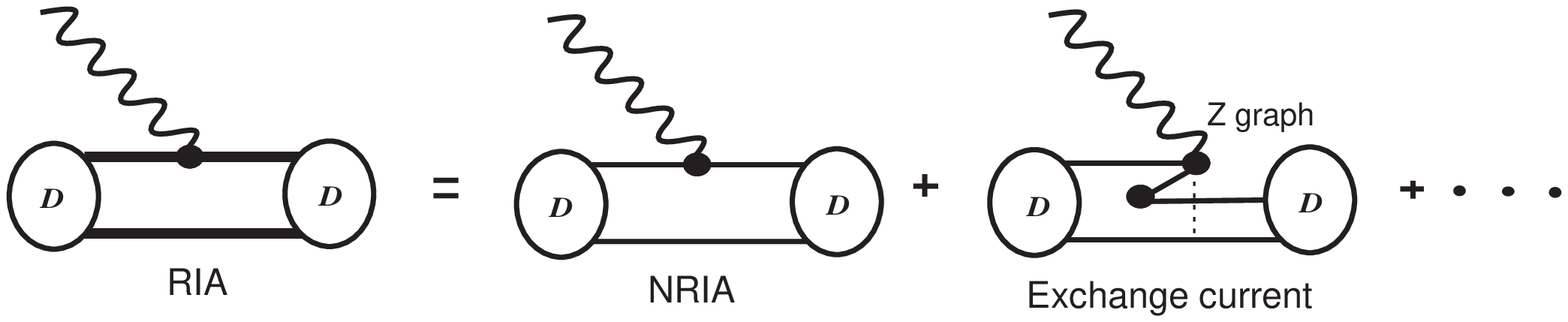}
\caption{Feynman diagrams for relativistic impulse approximation (RIA) and its non-relativistic decomposition.}
	 \label{fig:rd}
\end{minipage}
\end{figure}

%======================================================
Traditionally, nucleon-nucleon scattering and deuteron properties
such as electromagnetic properties are investigated in the non-relativistic framework.
For example, Gari and Hyuga investigated electromagnetic properties of the 
deuteron up to momentum transfer 2.4 GeV${}^2$~\cite{garihyuga,Gari:1976bd}. 
In order to explain the experimental data, 
they needed to introduce meson-exchange currents in addition to the impulse processes.
In 90's, Tamura et al., Blunden et al. and Wringa et al.~\cite{tamura,blunden,wiringa}
also investigated the electromagnetic properties of the 
deuteron in a non-relativistic framework.
In the non-relativistic framework, 
electron-deuteron scattering is expressed 
as shown in Fig.~\ref{fig:nrd},
where (a) is the impulse process and 
(b) is the contributions from the exchange currents. 
However, all non-relativistic treatments which are based on the impulse approximation with relativistic 
corrections and meson exchange currents are sensitive to the model of the meson exchange currents. 

In principle nucleons are described by the Dirac equation,
which is the relativistic equation.
In the relativistic framework,
we expect the following advantages:
\begin{itemize}
\item Relativistic kinematics are automatically included.
\item Some exchange currents are included through the $Z$-graph process.
\item Dynamics originated from the relativistic effect such as the LS-force, spin-spin interaction and the negative energy components
are naturally included.
\end{itemize}
In the relativistic framework, 
electron-deuteron scattering can be expressed 
as shown in Fig.~\ref{fig:rd}.
In the relativistic impulse approximation (RIA), each nucleon line has both negative and positive energy states,
in contrast with the non-relativistic approximation where each nucleon line contains only positive 
energy states.
The right hand side of Fig.~\ref{fig:rd} shows the decomposition of the RIA diagram 
into several diagrams in the non-relativistic reduction, where the diagrams are depicted in the chronological order.
One of those diagrams expresses the Z-graph whose nucleon line goes back as corresponding to the negative energy states. 
Therefore, we can interpret that the Z-graph part corresponds to the exchange current
in the non-relativistic approach.
It implies that a part of dynamics of the inclusion of the negative energy components
can be interpreted by the exchange currents.

In order to see the effects of the relativistic framework,
we treat the Bethe-Salpeter equation in the fully relativistic manner~\cite{bsgeneral}.
In 1989, Rupp and Tjon solved the Bethe-Salpeter equation (BSE) by the
covariant Graz II potential.
It is a simple covariantization of the Graz II potential
which was built by Mathelisch, Plessas and Schweiger
~\cite{Mathelitsch:mr,rupptjon:1990}.
Rupp and Tjon succeeded in reproducing ${}^1S_0$ and ${}^3S_1$-${}^3D_1$ 
phase shifts up to $600$ MeV
in their relativistic framework.
However, they did not include negative energy components.
For the investigation based on a fully relativistic framework, we should include the
negative energy states appropriately.

For the relativistic description of the deuteron,
we prepare a two-nucleon set including negative energy states
of $P$-wave in addition to the $^3S_1$ and $^3D_1$ states of positive energy. 
We solve the BSE by using the separable ansatz
including the negative energy states,
and determine the parameters to reproduce deuteron form factors.

The contents of this paper are as follows.
In section 2, we give a general formalism of the Bethe-Salpeter 
equation for the deuteron,
and provide a solution using the separable ansatz.
We discuss how to determine the parameters of 
separable potential in some detail.
In section 3, we show the relativistic kinematics for the elastic 
electron-deuteron scattering. 
In section 4, results and discussions are presented.
The final section is devoted to the conclusions of the present work.

%======================================================
\section{BS approach with a separable interaction}
%======================================================
\subsection{The BS equation}
Let us start with the BSE for the $NN$ $T$-matrix:
\begin{eqnarray}
\label{eq:T-matrix52}
T_{\alpha\beta,\delta\gamma}(P,p^\prime,p) &=&
V_{\alpha\beta,\delta\gamma}(P,p^\prime,p)
\nonumber
\\
+ 
 i \int\frac{d^4 k}{(2\pi)^4}&& \hspace{-1cm}
V_{\alpha\beta,\epsilon\lambda}(P,p^{\prime},k)
S^{(1)}_{\epsilon\eta}(P/2+k) S^{(1)}_{\lambda\rho}(P/2-k)
T_{\eta\rho,\delta\gamma}(P,k,p)\,,
\end{eqnarray}
where the Greek letters express spinor indices.
The $T$-matrix and the interaction kernel $V$
are the functions of the total momentum $P$,
and the relative momentum $p$ and $p^{\prime}$,
of the initial and final states. 
$S^{(1)}$ is the one nucleon free propagator. 
The bound state corresponds to a pole in the $T$-matrix at $P^2=M_B^2$,
where $M_B$ is the mass of the bound state:
\begin{eqnarray}
\label{T_and_Gamma}
T_{\alpha\beta,\gamma\delta}(P,k^\prime,k) =\frac{\Gamma_{\alpha\beta}(P,k')
\overline{\Gamma}_{\gamma\delta}(P,k)}{P^2-M^2}+R_{\alpha\beta,\gamma\delta}(P,k^\prime,k)\,.
\end{eqnarray}
Here $\Gamma_{\alpha\beta}$ is the vertex function of BSE, and
 $R_{\alpha\beta,\gamma\delta}$ is regular at $P^2=M_B^2$.
We can express the BS amplitude by the vertex function as :
\begin{eqnarray}
\label{eq:BSA-vertex}
\Phi_{\alpha\beta}(P,k)=
S^{(1)}_{\alpha\gamma}(\frac{P}{2}+k)
S^{(1)}_{\beta\delta}(\frac{P}{2}-k)\Gamma_{\gamma\delta}(P,k)\,,
\end{eqnarray}
and we obtain the equation for the BS amplitude
from Eqs.~(\ref{eq:T-matrix52}), (\ref{T_and_Gamma}) and (\ref{eq:BSA-vertex}) :
\begin{eqnarray}\label{eq:BSA-eq}
\Phi_{\alpha\beta}(P,k) =
iS^{(1)}_{\alpha\eta}(\frac{P}{2}+k)
S^{(1)}_{\beta\rho}(\frac{P}{2}-k) \int
\frac{d^4 k^{\prime\prime}}{(2\pi)^4}
V_{\eta\rho,\epsilon\lambda}(P,k,k^{\prime\prime})
\Phi_{\epsilon\lambda}(P,k^{\prime\prime})\,. 
\end{eqnarray}

\subsection{Solutions to the BSE}
After partial wave decompositions (see \cite{Bondarenko:2002zz}),
 the BS equation for the $T$-matrix in the center-of-mass frame is given by
\begin{eqnarray}\label{eq:sep01}
&&T_{\alpha\beta}(p_0^{\prime},|\bp^{\prime}|,p_0,|\bp|;s) =
V_{\alpha\beta}(p_0^{\prime},|\bp^{\prime}|,p_0,|\bp|;s)+
%\nonumber
\\
&&\frac{i}{2\pi^2}\,\int\,d k_0\,\mbf{k}^2\,d|\mbf{k}|\,
\sum\limits_{\gamma\delta}\,
V_{\alpha\gamma}(p_0^{\prime},|\bp^{\prime}|,k_0,|\mbf{k}|;s)\,
S_{\gamma\delta}(k_0,|\mbf{k}|;s)\,
T_{\delta\beta}(k_0,|\mbf{k}|,p_0,|\bp|;s)\,.
\nonumber
\end{eqnarray}
Here the Greek indices represent quantum numbers
($JLS\rho$),
and the summation takes over all partial waves.
When we include both positive and negative energy states,
we have eight states, namely,
$^3S_1^{+}$, $^3D_1^{+}$, $^3S_1^{-}$, $^3D_1^{-}$, $^1P_1^{e}$,
$^1P_1^{o}$, $^3P_1^{e}$ and $^3P_1^{o}$,
which are labeled as $1,\dots,8$.
Details of the structure of these amplitudes
are discussed in \cite{Bondarenko:2002zz}.
Therefore $T$, $V$ and $S$ are $8\times 8$ matrices 
where
$n$-th row and column corresponds to the $n$-th state.
In this eight dimensional basis,
the propagator $S_{\alpha\beta}$ is expressed by 
\begin{eqnarray}
\label{2nprgtr}
%\[
S=\left(
    \begin{array}{@{\,}cccccccc@{\,}}
    S_{+} & 0 & 0 & 0 & 0 & 0 & 0 & 0\\
    0      &S_{+}& 0 & 0 & 0 & 0 & 0 & 0\\
    0 & 0 & S_{-}  & 0 & 0 & 0 & 0 & 0\\
    0 & 0 & 0 & S_{-} & 0 & 0 & 0 & 0\\
    0 & 0 & 0 & 0 & S_{e} & 0 &S_{o}  &0 \\
    0 & 0 & 0 & 0 & 0 & S_{e} &0  &S_{o} \\
    0 & 0 & 0 & 0 & S_{o} & 0 & S_{e}  &0 \\
    0 & 0 & 0 & 0 & 0 & S_{o} &0  &S_{e}  
    \end{array}
\right)
%\]
\,,
\end{eqnarray}
where two nucleon propagators $S_{\alpha}$ are given as 
\begin{eqnarray}
S_{+} = 
\frac{1}{(\sqrt{s}/2-e_k+k_0+i\epsilon)}\frac{1}{(\sqrt{s}/2-e_k-k_0+i\epsilon)}\,,
\nonumber
\end{eqnarray}
\begin{eqnarray}
S_{-} = 
\frac{1}{(\sqrt{s}/2+e_k+k_0-i\epsilon)}\frac{1}{(\sqrt{s}/2+e_k-k_0-i\epsilon)}\,,
\nonumber
\end{eqnarray}
\begin{eqnarray}
S_{+-} =\frac{1}{(\sqrt{s}/2-e_k+k_0+i\epsilon)}
\frac{1}{(\sqrt{s}/2+e_k-k_0-i\epsilon)}\,,
\nonumber
\end{eqnarray}
\begin{eqnarray}
S_{-+} =\frac{1}{(\sqrt{s}/2+e_k+k_0-i\epsilon)}
\frac{1}{(\sqrt{s}/2-e_k-k_0+i\epsilon)}
\,,
\nonumber
\end{eqnarray}
\begin{eqnarray}
S_{e} = S_{ee}=S_{oo}=\frac{S_{+-}+S_{-+}}{2}\,,
\nonumber
\end{eqnarray}
\begin{eqnarray}
S_{o} = S_{eo}=S_{oe}=\frac{S_{+-}-S_{-+}}{2}\,.
\label{eq:tnps}
\end{eqnarray}

Now let us introduce the separable ansatz of rank $N$
in the following manner:
\begin{eqnarray}\label{eq:sep02}
V_{\alpha\beta}(p_0^{\prime},|\bp^{\prime}|,p_0,|\bp|;s)=
\sum\limits_{i,j=1}^{N}\,
\lambda_{ij}\,g_i^{(\alpha)}(p_0^{\prime},|\bp^{\prime}|)\,
g_j^{(\beta)}(p_0,|\bp|),\quad \lambda_{ij}=\lambda_{ji},
\end{eqnarray}
where we assume that $\lambda_{ij}$ is symmetric
under the interchange of $i,j$.
Then the $T$-matrix is also given in a separable form as
\begin{eqnarray}
T_{\alpha\beta}(p_0^{\prime},|\bp^{\prime}|,p_0,|\bp|;s)=
\sum\limits_{i,j=1}^{N}\,
\tau_{ij}\,g_i^{(\alpha)}(p_0^{\prime},|\bp^{\prime}|)\,
g_j^{(\beta)}(p_0,|\bp|)\,,
\end{eqnarray}
 and  
\begin{eqnarray}
(\tau^{-1}(s))_{ij}=(\lambda^{-1})_{ij}-H_{ij}(s)\,.
\end{eqnarray}
Here $H_{ij}(s)$ is defined by  
\begin{eqnarray}
H_{ij}(s)
\hspace{-0.3cm} 
&=&
\hspace{-0.3cm}
\frac{i}{2\pi^2}
\hspace{-0.2cm}
\sum\limits_{LS\rho\rho^{\prime}} \int
\hspace{-0.1cm}
d k_0\,\mbf{k}^2\,
d|\mbf{k}|\,S_{\rho\rho^{\prime}}(k_0,|\mbf{k}|;s)
\,g_i^{(JLS\rho)}(k_0,|\mbf{k}|)\,
g_j^{(JLS\rho^{\prime})}(k_0,|\mbf{k}|).
\end{eqnarray}
Then, the radial part of the BS amplitude
can be written as
\begin{eqnarray}
 \phi_{JLS\rho}(p_0,|\bp|) &=&\sum\limits_{\rho^{\prime}}
S_{\rho\rho^{\prime}}(p_0,|\bp|)\,g_{JLS\rho}(p_0,|\bp|)
\nonumber
\\
&=&\sum\limits_{\rho^{\prime}}\sum_{i,j=1}^{N}\,
S_{\rho\rho^{\prime}}(p_0,|\bp|;s)
\lambda_{ij} g_{i}^{(JLS\rho^{\prime})}(p_0,|\bp|) c_j(s)\,.
\label{eq:sep06}
\end{eqnarray}
where $c_i(s)$ satisfy the following equation:
\begin{eqnarray}
\label{eq:sep05}
c_i(s)&-&\sum\limits_{k,j=1}^{N} H_{ik}(s)\lambda_{kj}c_j(s) = 0,
\end{eqnarray}
Using $\phi_{JLS\rho}(p_0,|\bp|)$, 
%is the radial part function of the Eq.~(\ref{ampmat}),
we can obtain the general form of 
the BS amplitude (see \cite{Bondarenko:2002zz}).

Because $^3P_1$ spin-angular part has a complex form (see \cite{Bondarenko:2002zz}),
we include $^1P_1^{o}$ and $^1P_1^{e}$ 
but ignore $^3P_1^{o}$ and $^3P_1^{e}$ among negative energy components
for simplicity in actual calculation.
We also ignore $^3S_1^{-}$, $^3D_1^{-}$ states,
because we expect that the contribution from those two states is second order
in the non-relativistic expansion (in powers of velocity)
and is small.

\subsection{Rank IV separable ansatz}
For a relativistic description of the deuteron 
when we include both the positive and negative energy components
in the BS amplitude,
we extend the covariant Graz-II interaction with including $P$-wave components. 
For this purpose, we adopt a rank IV separable ansatz.
The Graz-II interaction is a rank III separable potential suggested by Mathelitsch et al. in the
non-relativistic framework~\cite{Mathelitsch:mr}.
Rupp made a covariant Graz-II potential by simple covariantization~\cite{rupptjon:1990}.
In the covariant Graz-II interaction,
they considered only positive energy states, $^3S_1^{+}$ and $^3D_1^{+}$.
Here in this work we build a rank IV separable potential ($N=4$) by improving the rank III
covariant Graz-II interaction with the inclusion of the negative energy states.
We use the following forms for the functions $g_i^{\alpha}$, 
\begin{eqnarray}
&&g_{1}^{^3S_1^+}(p_0,|\bp|)=\frac{1-\gamma_{1}(p_0^2-{\bp}^2)}
{(p_0^2-{\bp}^{2}-\beta_{11}^{2}+i\epsilon)^{2}},
\nonumber\\
%\label{gfactors}\\
&&g_{2}^{^3S_1^+}(p_0,|\bp|)=-\frac{(p_0^2-{\bp}^2)}
{(p_0^2-{\bp}^2-\beta_{12}^2+i\epsilon)^{2}},
\nonumber \\
&&g_{3}^{^3D_1^+}(p_0,|\bp|)=\frac{(p_0^2-{\bp}^{2})
(1-\gamma_{2}(p_0^2-{\bp}^{2}))}
{(p_0^2-{\bp}^{2}-\beta_{21}^{2}+i\epsilon)(p_0^2-{\bp}^{2}-\beta_{22}^{2}+i\epsilon)^{2}},
\label{gfactors} \\
\nonumber
&&g_{4}^{^1P_1^e}(p_0,|\bp|)=\frac{|\bp|}
{(p_0^2-{\bp}^2-\beta_{3}^2+i\epsilon)^{2}} ,\\\nonumber
&&g_{4}^{^1P_1^o}(p_0,|\bp|)=\gamma_{3}\frac{ p_0}{m}\frac{|\bp|}
{(p_0^2-{\bp}^2-\beta_{3}^2+i\epsilon)^{2}}\,,
\\
&&g_{3,4}^{^3S_1^+}(p_0,|\bp|)=g_{1,2,4}^{^3D_1^+}(p_0,|\bp|)
=g_{1,2,3}^{^1P_1^{e,o}}(p_0,|\bp|)= 0 \,,
\nonumber
\\
&&g_{1,2,3,4}^{^3S_1^-}(p_0,|\bp|)=g_{1,2,3,4}^{^3D_1^-}(p_0,|\bp|)
=g_{1,2,3,4}^{^3P_1^{e,o}}(p_0,|\bp|)= 0 \,.
\nonumber
\end{eqnarray}
For $^3S_1^{+}$ and $^3D_1^{+}$, we use the same type of function 
as the covariant Graz-II Interaction.
What is new here is the inclusion of functions for the $^1P_1^{e,o}$-waves.
The last line reflects our assumption that 
we ignore $^3P_1^{o}$, $^3P_1^{e}$, $^3S_1^{-}$ and $^3D_1^{-}$.

Having the ansatz Eq.~(\ref{gfactors}),
the solution Eq.~(\ref{eq:sep06}) to the BSE can be written as
\begin{eqnarray}\label{vert-graz}
\phi_{^3S_1^+}(p_0,|\bp|) &=&
(c_1 \lambda_{11}+c_2 \lambda_{12}+c_3 \lambda_{13}+c_4 \lambda_{14})
S_{+} g_1^{^3S_1^+}(p_0,|\bp|)+ \\
&& (c_1 \lambda_{12}+c_2 \lambda_{22}+c_3 \lambda_{23}+c_4 \lambda_{24})
S_{+} g_2^{^3S_1^+}(p_0,|\bp|), \nonumber\\\nonumber
\phi_{^3D_1^+}(p_0,|\bp|) &=&
(c_1 \lambda_{13}+c_2 \lambda_{23}+c_3\lambda_{33}+c_4 \lambda_{34})
S_{+} g_3^{^3D_1^+}(p_0,|\bp|),\\\nonumber
\phi_{^1P_1^e}(p_0,|\bp|) &=&
(c_1 \lambda_{14}+c_2 \lambda_{24}+c_3\lambda_{34}+c_4 \lambda_{44})
(S_{e} g_4^{^1P_1^e}(p_0,|\bp|)+S_{o} g_4^{^1P_1^o}(p_0,|\bp|)),\\\nonumber
\phi_{^1P_1^o}(p_0,|\bp|) &=&
(c_1 \lambda_{14}+c_2 \lambda_{24}+c_3\lambda_{34}+c_4 \lambda_{44})
(S_{e} g_4^{^1P_1^o}(p_0,|\bp|)+S_{o} g_4^{^1P_1^e}(p_0,|\bp|))\,,
\nonumber
\\
\phi_{^3S_1^-}(p_0,|\bp|) &=&\phi_{^3D_1^-}(p_0,|\bp|)=\phi_{^3P_1^o}(p_0,|\bp|)=\phi_{^3P_1^e}(p_0,|\bp|)=0\,,
\nonumber
\end{eqnarray}
where we have used the facts that $\lambda_{ij}=\lambda_{ji}$,
and that $\phi_{^1P_1^e}$ is even and that $\phi_{^1P_1^o}$ is odd
under $p_0 \rightarrow -p_0$.

\subsection{Computations with negative energy components \label{section:cwnec}}
When we use the rank IV separable potential,
Eq.~(\ref{eq:sep05}) takes the following form
\begin{eqnarray}
c_1-\sum^3_{k, j=1}H_{1k}\lambda_{kj}c_j&=&c_4(H_{11}\lambda_{14}+H_{12}\lambda_{24})\,,\nonumber\\
c_2-\sum^3_{k, j=1}H_{2k}\lambda_{kj}c_j&=&c_4(H_{21}\lambda_{14}+H_{22}\lambda_{24})\,,\nonumber\\
c_3-\sum^3_{k, j=1}H_{3k}\lambda_{kj}c_j&=&c_4(H_{33}\lambda_{34})\,,\nonumber\\
c_4-\sum^3_{k, j=1}H_{4k}\lambda_{kj}c_j&=&c_4(H_{44}\lambda_{44})\,.
\end{eqnarray}
In these equations,
$\lambda_{j4}$, $H_{4j}(j=1,2,3,4)$ and $c_4$ appear 
due to the inclusion of $^1P^e_1$ and $^1P^o_1$-states.
Since the fourth component in the rank IV ansatz is introduced for the first time
in this work, we would like to discuss in some detail the quantities carrying the fourth index.
First, we note that $H_{4j}(j=1,2,3)=0$ due to the choice of the $g$-functions
as given in Eqs.~(\ref{gfactors}).
On the other hand, $H_{44}$ can take a finite value.
We would like to discuss $H_{44}$ in detail.

The evaluation of $H_{44}$ contains a $k_0$-integral,
and the result is affected by the location of poles of the $g$-functions
and propagator $S_{\alpha}$ as defined by Eq.~(\ref{eq:tnps}).    
For deuteron problems, 
we take the location of poles of $g^{^1P_1^e}$ and $g^{^1P_1^o}$ in 
the same side of the complex $k_0$-plane as $g^{^3S_1^+}$.
As we will discuss in detail later, $P$-wave part influence on the form factors and tensor polarizations
through $H_{44}$, $c_j (j=1,2,3,4)$ and $\lambda_{4j}(j=1,2,3,4)$.
This is an important contribution in order to improve the agreement with the 
experimental data
through the negative energy components in bound state problems.

\begin{figure}[] 
\begin{center}
\begin{minipage}{12cm}
\includegraphics[origin=c,width=6cm,height=4cm]{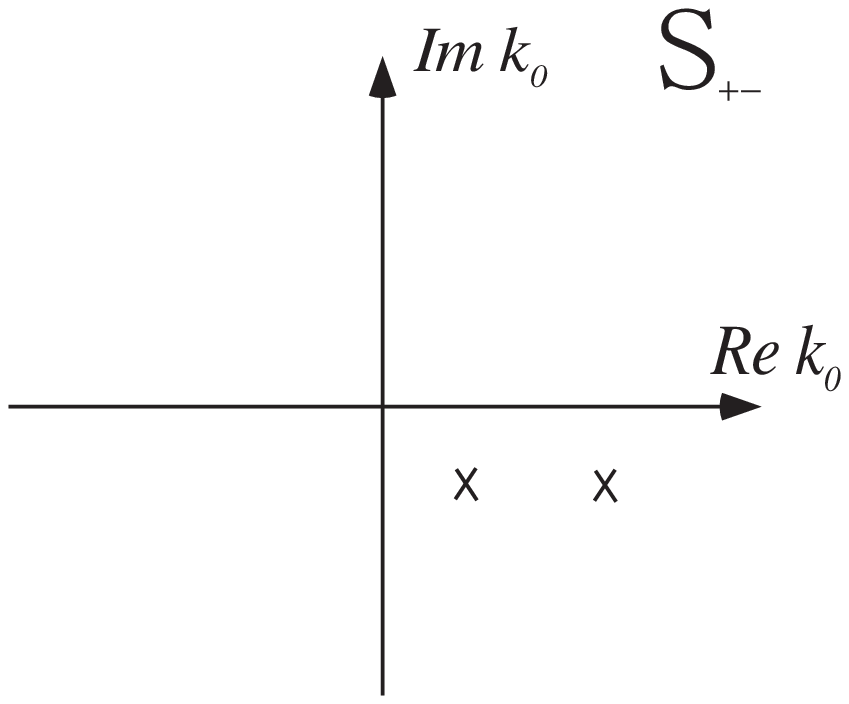}
\includegraphics[origin=c,width=7.5cm,height=5cm]{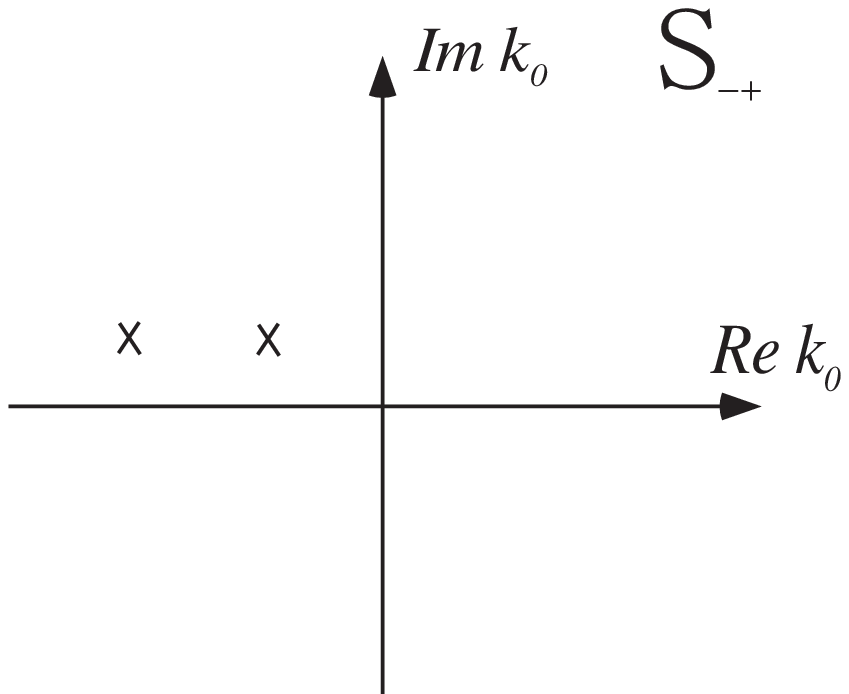}
\end{minipage}
\end{center}
\caption{Location of poles of $S_{+-}$ and $S_{-+}$}
       \label{fig:cps}
\end{figure}

\begin{figure}[ht]
\centering
\begin{minipage}{14cm}
\includegraphics[origin=c,angle=0,width=7cm,height=6cm]{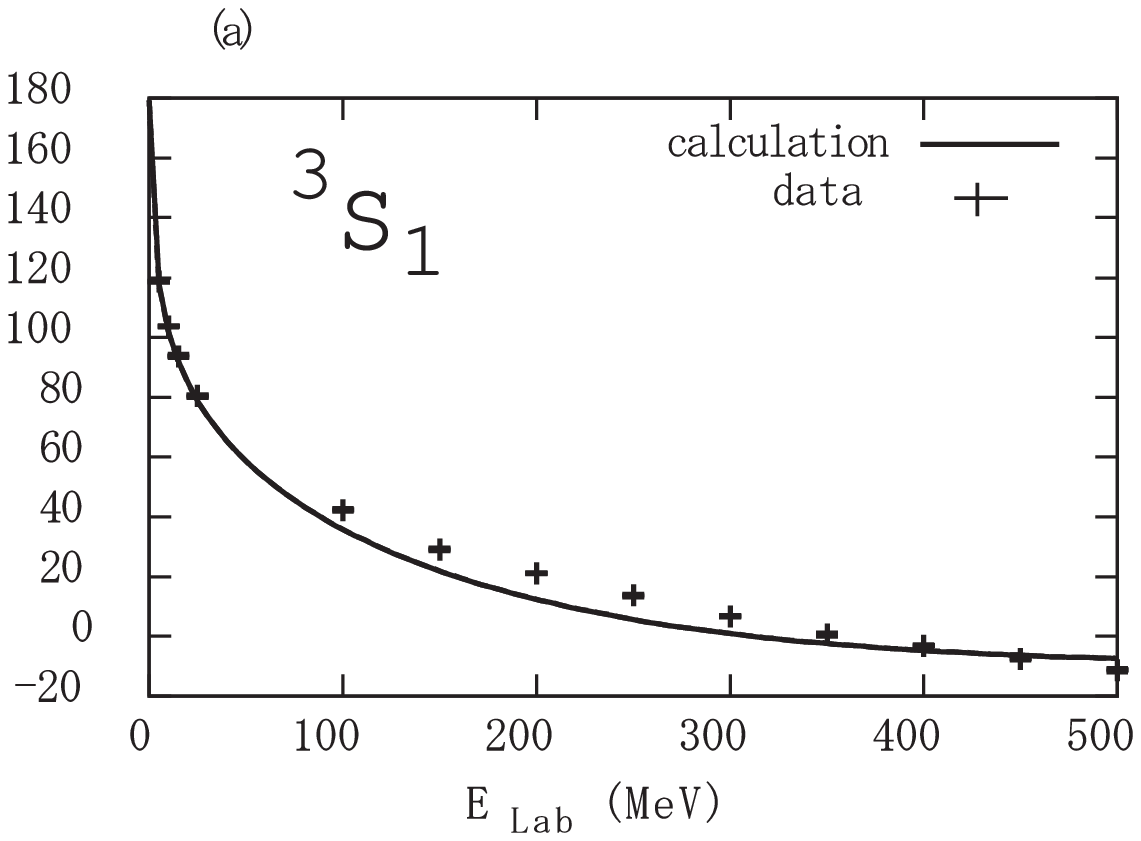}
\hspace{-0.5cm}
\includegraphics[origin=c,angle=0,width=7cm,height=6cm]{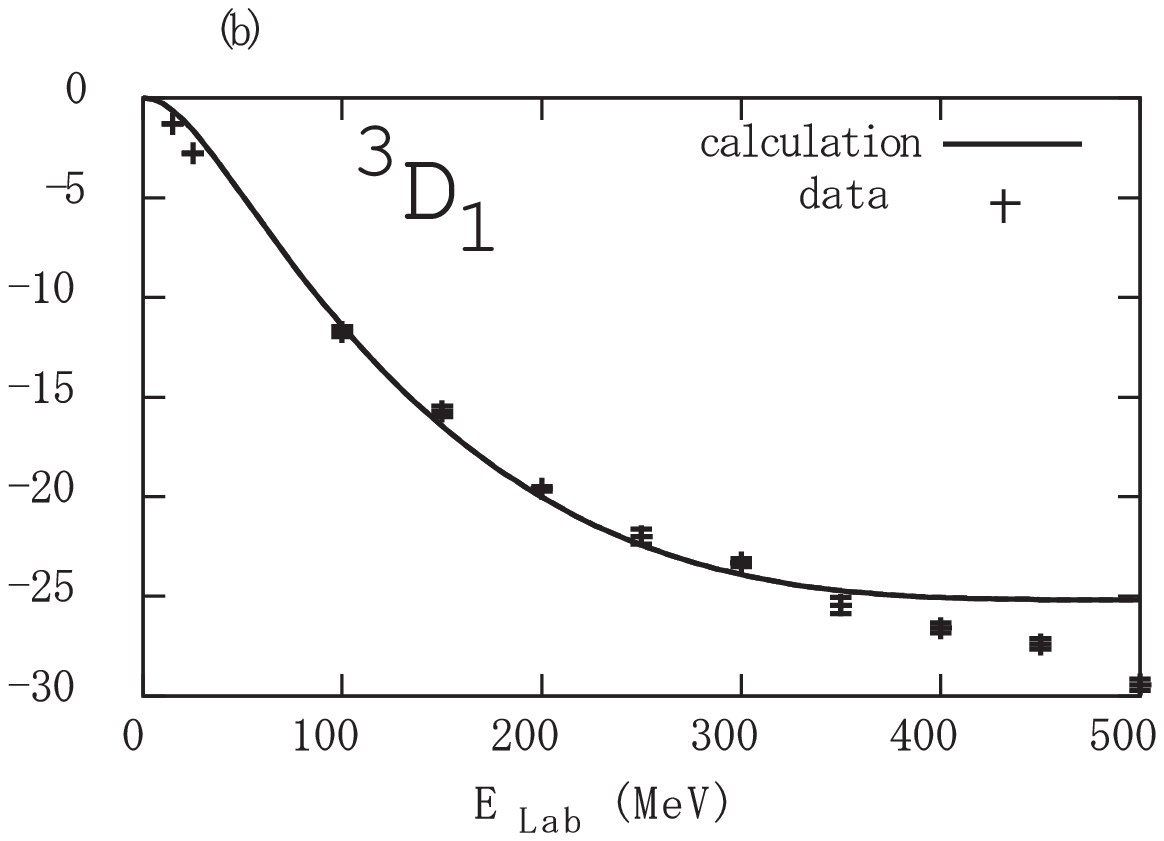}
\caption{Phase shifts of $^3S_1$ (left) and $^3D_1$ (right) channels. 
Data are taken from SAID program (http://gwdac.phys.gwu.edu).
Calculational results (solid line) are performed with the $D$-wave ratio $4\%$
and with the negative energy $P$-wave ratio $0$, $-1$ and $-3\%$
(for negative values of the $P$-wave ratio, see the discussion in Appendix C).
Results with three $P$-wave ratios are almost the same within the resolution of the figures.}
          \label{fig:3s13d1d4p013}
\end{minipage}          
\end{figure}

However, if we adopt the same prescription for scattering problems,
a finite $H_{44}$ causes a serious problem, which change the phase shift drastically,
even if the rate of $P$-wave is very small.
Therefore, we try to solve this problem in the following way. 
For $H_{44}$, we have the following expression:
\begin{eqnarray}
H_{44}&=&
\frac{i}{2\pi^2} \int\,d k_0\,\mbf{k}^2\,       
d|\mbf{k}|\,
\left(
S_{e}\,((g_4^{^1P_1^e})^2+(g_4^{^1P_1^o})^2)
+2S_{o}\,(g_4^{^1P_1^e}\,g_4^{^1P_1^o})
\right)
\nonumber 
\\
&=&
\frac{i}{2\pi^2} \int\,d k_0\,\mbf{k}^2\,
d|\mbf{k}|\,
\left(
S_{+-}\,
(g^{+-})^2
+S_{-+}\,(g^{-+})^2
\right)\,,
\label{eq:h44}
\end{eqnarray}
where
\begin{eqnarray}
g^{+-}\equiv g_4^{^1P_1^e}+g_4^{^1P_1^o}\,,
\nonumber
\end{eqnarray}
\begin{eqnarray}
g^{-+}\equiv g_4^{^1P_1^e}-g_4^{^1P_1^o}\,.
\nonumber
\end{eqnarray}
At first sight, Eq.~(\ref{eq:h44}) takes a finite value.
However, by locating the poles of $g^{+-}$ in the upper side of the complex $k_0$-plane
as $S^{+-}$, and the poles of $g^{-+}$ lower as $S^{-+}$ (see Fig.~\ref{fig:cps}),
Eq.~(\ref{eq:h44}) becomes zero.      
This prescription to make the vanishing contribution for $H_{44}$
turns out to be crucially important in order to reproduce
the experimental phase shifts (see Fig.~\ref{fig:3s13d1d4p013}).
Physically, this means that we neglect the direct contribution from negative energy components
for scattering problems.
For the present separable potential of $g$-functions,
we have realized this by locating the pole in the appropriate manner.
As we discuss in Appendix \ref{sapole}, 
the similar situation can be realized for static separable potential,
which justifies the present prescription.

By assuming different pole location of the $g$-function for the bound state and 
scattering problems,
we are able to reproduce reasonable results for both quantities.
The physical ground of this condition is related to the fact
that a finite value of $H_{44}$ allows appearance of negative energy
states in intermediate states.
For a bound state, negative energy components are naturally contained in
the relativistic formalism, and therefore, the mixture of negative  
energy state is allowed.
In contrast, for scattering problems such mixture is not allowed, since
scattering states are expressed as definite energy states as in the
asymptotic states.
Therefore, $H_{44}$ must vanish for scattering problems.
In this way, we will be able to obtain good results for both 
scattering and bound state (deuteron) problems simultaneously. 

In actual calculations,
we express the $\lambda_{i4}$ parameters in terms of a single $u_4$ parameters by
\begin{eqnarray}
\label{eq:lamp}
\lambda_{14}=-\sqrt{\lambda_{11}}u_{4},\,
\lambda_{24}=\sqrt{\lambda_{22}}u_{4},\,
\lambda_{34}=\sqrt{\lambda_{33}}u_{4},\,
\lambda_{44}=u_{4}^2\,.
\end{eqnarray}
We try to find good parameter sets to reproduce the form factors, tensor polarizations
and phase shifts.

\section{Elastic Electron-Deuteron Scattering}
\subsection{Relativistic Kinematics\label{RK}}
The differential cross section for unpolarized elastic
electron-deuteron scattering in the one-photon-exchange
approximation 
is given by
\begin{eqnarray}
\frac{d\sigma}{d\Omega_{\el}^{\prime}} =
\Bigl(\frac{d\sigma}{d\Omega_{\el}^{\prime}}\Bigr)_{\rm Mott}
\Bigl[A(q^2)+B(q^2)\tan^2{\frac{\theta_{\el}}{2}}\Bigr]\,.
\label{cross}
\end{eqnarray} 
Here $\Bigl(\frac{d\sigma}{d\Omega_{\el}^{\prime}}\Bigr)_{\rm Mott}$
is the Mott cross section given by 
\begin{eqnarray}
\Bigl(\frac{d\sigma}{d\Omega_{\el}^{\prime}}\Bigr)_{\rm Mott}=
\frac{\alpha^2
\cos^2{\theta_{\el}/2}}{4E_{\el}^2(1+2E_{\el}/M\sin^4{\theta_{\el}/2})},
\end{eqnarray}
where $\theta_{\el}$ is the electron scattering angle, 
$M$ the deuteron mass
and $E_e$ the incident electron energy.
The functions $A(q^2)$ and $B(q^2)$ are the deuteron structure functions 
which can be related to the form factors by
\begin{eqnarray}
&&A(q^2)=F_{{\rm C}}^2(q^2)+\frac{8}{9}\eta^2F_{\rm Q}^2(q^2)+
\frac{2}{3}\eta F_{\rm M}^2(q^2),
\nonumber\\
&&B(q^2)=\frac{4}{3}\eta(1+\eta)F_{\rm M}^2(q^2),
\label{structf}
\end{eqnarray}
where $\eta=-{q^2}/{4M^2} = Q^2/4M^2$.
The electric $F_{\rm C}(q^2)$, the magnetic $F_{\rm M}(q^2)$ 
and the quadrupole $F_{\rm Q}(q^2)$ 
form factors are normalized as
\begin{eqnarray}
F_{\rm C}(0)=1, \quad 
F_{\rm M}(0)=\mu_{\rm D}\frac{M}{m}, \quad 
F_{\rm Q}(0)=M^2 \, Q_{\rm D}\,,
\label
{normf}
\end{eqnarray} 
where $m$ is the mass of the nucleon,
$\mu_{\rm D}$ the magnetic moment 
and $Q_{\rm D}$ the quadrupole moment of the deuteron. 
The tensor polarization components of the final deuteron can be written 
through the deuteron form factors as follows:
\begin{eqnarray}
&& t_{20}\ \bigl[A+B\tan^2{\frac{\theta_{\el}}{2}}\bigr]=
-\frac{1}{\sqrt{2}}\bigl[\frac{8}{3}\eta F_{\rm C}F_{\rm Q}+
\frac{8}{9}\eta^2F_{\rm
  Q}^2+\frac{1}{3}\eta(1+2(1+\eta)\tan^2{\frac{\theta_{\el}}{2}})F_{\rm M}^2\bigr],
\nonumber\\
&& t_{21}\ \bigl[A+B\tan^2{\frac{\theta_{\el}}{2}}\bigr]=
\frac{2}{\sqrt{3}}\eta(\eta+\eta^2\sin^2{\frac{\theta_{\el}}{2}})^{1/2}
F_{\rm M} F_{\rm Q} \sec{\frac{\theta_{\el}}{2}},
\label{tensormom}\\
&& t_{22}\ \bigl[A+B\tan^2{\frac{\theta_{\el}}{2}}\bigr]=
-\frac{1}{2\sqrt{3}}\eta F_{\rm M}^2. 
\nonumber
\label{eq:tensorp01}
\end{eqnarray}

We can obtain Eq.~(\ref{cross})
from the following amplitude
\begin{eqnarray}
M_{\rm fi}=i e^2{\bar u_{m^{\prime}}(l^{\prime})}
\gamma^{\mu}u_{m}(l)\;\frac{1}{q^2} \;\langle D^{\prime}{\cal
M}^{\prime} | J_{\mu} | D {\cal M} \rangle,
\label{eqn:M}
\end{eqnarray} 
where $u_{m}(l)$ is the free electron spinor with 4-momentum $l$
and the spin projection $m$,
$q=l-l^{\prime}=P^{\prime}-P$ the 4-momentum transfer
and $P\,(P^{\prime})$ the initial (final) deuteron momentum. 
$|D{\cal M}\rangle$ is the deuteron state with an angular momentum
projection ${\cal M}$ 
and $J_{\mu}$ is the electromagnetic current operator.

The deuteron current matrix element is parameterized in
the following way 
\begin{eqnarray}
\langle D^\prime {\cal M}^\prime |J_\mu|D{\cal M}\rangle= &-&e\;
\xi^*_{\alpha\;{\cal M}^\prime}(P^{\prime})
\Biggl[ (P^\prime+P)_{\mu}
\Bigl(g^{\alpha\beta}F_1(q^2)-\frac{q^\alpha
q^\beta}{2M^2}F_2(q^2)\Bigr)
\nonumber\\
&-& (q^{\alpha} g^\beta_\mu - q^{\beta} g^\alpha_\mu)G_1(q^2)
\Biggr]
\;\xi_{\beta\;{\cal M}}(P)\; 
\,, 
\label{deuteronc}
\end{eqnarray} 
where $\xi_{\cal M}(P)$
and $\xi^{*}_{{\cal M}^\prime}(P^\prime)$ are the polarization
4-vectors of the initial and final deuteron.
Form factors $F_{1,2}(q^2)$, $G_1(q^2)$ are related to $F_{\rm
C}(q^2)$, $F_{\rm Q}(q^2)$ and $F_{\rm M}(q^2)$ by the equations
\begin{eqnarray}
F_{\rm C} = F_1 + \frac{2}{3}\eta \bigl[F_1 + (1+\eta)F_2 - G_1
\bigr]\,,
F_{\rm Q} = F_1 + (1+\eta)F_2 - G_1\,,
F_{\rm M} = G_1\,. 
\label{ffconnect}
\end{eqnarray} 
The normalization condition for the deuteron current matrix element
is given by
\begin{eqnarray}
\lim_{q^2\to 0} \langle D^\prime {\cal M}^\prime |J_\mu|D{\cal
M}\rangle = 2eP_{\mu}\; \delta_{{\cal M}{\cal M}^{\prime}}\,.
\nonumber
\end{eqnarray}
Equivalently, in terms of the BS amplitude:
\begin{eqnarray}
\int \frac{d^4k}{(2\pi)^4}
\overline\Gamma(P,k)
\frac{\partial}{\partial {P_{\mu}}} \left\{
{S}^{(1)}(P/2+k){S}^{(1)}(P/2-k)\right\} \Gamma(P,k) =
-2P_{\mu} \,.
\label{normalization2}
\end{eqnarray}

In order to calculate the deuteron form factors,
let us choose the laboratory frame (deuteron at rest). 
In the laboratory frame the relevant momentum variables take 
the following form (the $z$-axis is along the photon momentum):
\begin{eqnarray}
&&P=(M,{\bf 0}), \quad P^{\prime}=P+q=(M(1+2\eta),0,0,2 M
\sqrt{\eta}\sqrt{1+\eta}),
\nonumber\\
&&q=(2M\eta,0,0,2M\sqrt{\eta}\sqrt{1+\eta}),
\label{vlab}
\end{eqnarray}
\begin{eqnarray}
&&\xi_{{\cal M}=+1}(P) = \xi_{{\cal M}=+1}(P^{\prime})=
-\frac{1}{\sqrt{2}} (0,1,i,0),
\nonumber\\
&&\xi_{{\cal M}=-1}(P) = \xi_{{\cal M}=-1}(P^{\prime})=
\frac{1}{\sqrt{2}} (0,1,-i,0), \nonumber\\
&&\xi_{{\cal M}=0}(P) = (0,0,0,1),\quad \xi_{{\cal
M}=0}(P^{\prime})= (2\sqrt{\eta}\sqrt{1+\eta},0,0,1+2\eta).
\label{xibreit}
\end{eqnarray}
From Eqs.~(\ref{vlab}), (\ref{xibreit}) and Eq.~(\ref{deuteronc}),
we obtain:
\begin{eqnarray}
&&\langle {\cal M}^{\prime} | J_{0} | {\cal M} \rangle =\; 2Me\;
(1+\eta)\; \Bigl\{ F_1 \delta_{{\cal M} {\cal M}^{\prime}} + 2\eta
\bigl[F_1 + (1+\eta)F_2 - G_1 \bigr] \delta_{{\cal M}^{\prime} 0}
\delta_{{\cal M} 0}\Bigr\}, \nonumber\\
&&\langle {\cal M}^{\prime} | J_{x} | {\cal M} \rangle =\;
\frac{2Me}{\sqrt{2}}\; \sqrt{\eta}\;\sqrt{1+\eta}\; G_1\; \Bigl\{
\delta_{{\cal M}^{\prime} {\cal M}+1} - \delta_{{\cal M}^{\prime}
{\cal M}-1} \Bigr\}. 
\label{emcmatel}
\end{eqnarray}

\subsection{Deuteron currents and form factors}
\begin{figure}
\centering
\begin{minipage}{8cm}
\centering
\includegraphics[width=8cm]{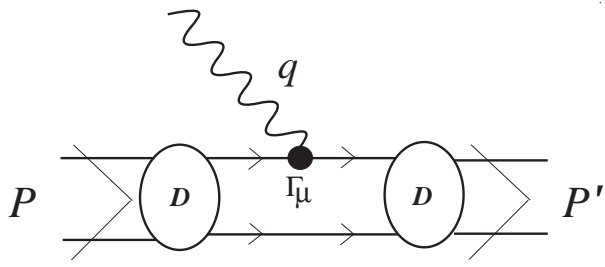}
\caption{A schematic diagram for the electromagnetic process of the deuteron.}
	 \label{fig:dc}
\end{minipage}
\end{figure}
In the relativistic impulse approximation,
a deuteron current matrix element can be written as
\begin{eqnarray}
\label{eq:RIA01}
\langle D^{\prime}\mathcal{M}^{\prime} | J_{\mu} | D \mathcal{M}
\rangle
&=& i {\rm e} \int \frac{d^4p}{(2\pi)^4} {\textrm Tr} \left[
\overline{\Phi}_{\mathcal{M}^{\prime}}(P^{\prime},p^{\prime})
\Gamma^{(p+n)}_{\mu}(q) \Phi_\mathcal{M}(P,p)\,(S^{(2)T}(q_2))^{-1}
\right],
\nonumber
\\
\Gamma_{\mu}(q) &=& \gamma_{\mu} F_1(q^2)
-\frac{\gamma_{\mu} {\hat q} - {\hat q} \gamma_{\mu}}{4m} F_2(q^2),
\label{eq:riadc}
\end{eqnarray}
where $\Phi_{\cal M}(P,p)$ is the BS amplitude of the deuteron,
$P^{\prime}=P+q$ and $p^{\prime}=p+q/2$,
where
$q$ is the momentum transfer.
$\Gamma_{\mu}(q)$ is the vertex of $\gamma NN$ interaction,
$F_1$ and $F_2$ are the Dirac and Pauli form factors of the nucleon,
and the sum over the proton and neutron is taken.
A schematic diagrams is shown in Fig.~\ref{fig:dc}.  

The resulting expressions for the deuteron current matrix element
can be written as
\begin{eqnarray}
&& \langle D^{\prime}{\cal M}^{\prime} | J^{RIA}_{\mu} | D {\cal
M} \rangle ={\cal I}_{1\;\mu}^{{\cal M}^{\prime}{\cal
M}}(q^2)\;F_1(q^2)+ {\cal I}_{2\;\mu}^{{\cal
M}^{\prime}{\cal M}}(q^2)\;F_2(q^2),
\label{fffd}\\
&& {\cal I}_{1,2\;\mu}^{{\cal M}^{\prime}{\cal M}}(q^2)=ie \int d
k_0\;|\bk|^2\;d|\bk|\;d(\cos{\theta})
\nonumber\\
&&
\times
\hspace*{-.5cm}\sum_{L^{\prime}S^\prime\rho^\prime,LS\rho}\hspace*{-.5cm}
\phi_{_{1L^{\prime}S^\prime\rho}}(k_0^{\prime},|\bk^{\prime}|)
\phi_{_{1LS\rho}}(k_0,|\bk|)\;I^{L^{\prime},L}_{1,2\;{\cal
M^{\prime}}{\cal M}\;\mu} (k_0,|\bk|,\cos{\theta},q^2),
\nonumber
\end{eqnarray} 
where the function
$I^{L^{\prime},L}_{1,2\;{\cal M^{\prime}}{\cal M}\;\mu}
(k_0,|\bk|,\cos{\theta},q^2)$ is obtained by taking the trace in the
$\gamma$ matrix space of Eq.~(\ref{eq:riadc})
and the substitution of the scalar products into Eq.~(\ref{eq:RIA01}).

In Eq.~(\ref{fffd}),
the radial part of the BS amplitude for the final state deuteron
$\phi_{1L^{\prime}S^\prime\varrho^\prime}(k_0^{\prime},|\bk^{\prime}|)$
depends on the momentum variable $k^{\prime}$
in the laboratory frame. 
The momenta of the initial deuteron ($P$)
and of the final deuteron ($P'$) in the laboratory frame 
are related 
by the Lorentz transformation:
\begin{eqnarray}
P^{\prime}= {\cal L} P = {\cal L} (M,{\bf
0}), \quad k^{\prime} = {\cal L} k,
\label{transfv}\end{eqnarray} where the Lorentz transformation
matrix ${\cal L}$ is of the form:
\begin{eqnarray}
{\cal L} = \left(
\begin{array}{cccc}
1+2\eta & 0 & 0 & 2\sqrt{\eta}\sqrt{1+\eta} \\
0 & 1 & 0 & 0 \\
0 & 0 & 1 & 0 \\
2\sqrt{\eta}\sqrt{1+\eta} & 0 & 0 & 1+2\eta
\end{array}
\right)\,. 
\label{mtrtransf}
\end{eqnarray}

The components of the 4-vector of the final state
$k^{\prime}$ are expressed by 
$k^{\prime}  =
(k_0^{\prime},k_x^{\prime},k_y^{\prime},k_z^{\prime})$, and
$|{\bk}^{\prime}|=\sqrt{k_x^{\prime\,2} +
k_y^{\prime\,2} + k_z^{\prime\,2}}$. 
Using Eq.~(\ref{vlab}), (\ref{transfv}) and (\ref{mtrtransf}) 
we obtain
\begin{eqnarray}
k_0^{\prime} &=& (1+2\eta)k_0 - 2\sqrt{\eta}\sqrt{1+\eta}k_z -
M\eta,
\nonumber\\
k_x^{\prime} &=& k_x, \quad k_y^{\prime} = k_y,\nonumber \\
k_z^{\prime} &=& (1+2\eta)k_z - 2\sqrt{\eta}\sqrt{1+\eta}k_0 +
M\sqrt{\eta}\sqrt{1+\eta}, 
\label{pprimecm}
\end{eqnarray} 
where $k_0, k_x, k_y, k_z$ are the components of the 4-vector $k$ of the
initial state.

In order to calculate the deuteron form factors, 
we need to know three matrix elements with different
total angular momentum projections and current components,
for example, $\langle 0 | J_{0} | 0 \rangle $,
$\langle 1 | J_{0} | 1 \rangle $ and $\langle 1 | J_{x} | 0 \rangle $.
Finally we obtain the following equations.
\begin{eqnarray}
F_C&=&\frac{1}{2M}\frac{\langle P^{\prime}\mathcal{M}^{\prime}=0 | J_{0} | P \mathcal{M}=0\rangle
+2\langle P^{\prime}\mathcal{M}^{\prime}=+1 | J_{0} | P \mathcal{M}=+1\rangle
 }
{3(1+\eta)}
\,,
\nonumber\\ 
F_M&=&\frac{1}{M\sqrt{2}}\frac{\langle P^{\prime}\mathcal{M}^{\prime}=+1 | J_{x} | P \mathcal{M}=0\rangle}
{\sqrt{\eta}\sqrt{1+\eta}}
\,,
\nonumber\\
F_Q&=&\frac{1}{2M}\frac{\langle P^{\prime}\mathcal{M}^{\prime}=0 | J_{0} | P \mathcal{M}=0\rangle
-\langle P^{\prime}\mathcal{M}^{\prime}=+1 | J_{0} | P \mathcal{M}=+1\rangle
 }
{{\eta}\sqrt{1+\eta}}
\,.
\label{eq:nngs}
\end{eqnarray}
We can obtain tensor polarization components by inserting these equations into 
Eqs.~(\ref{eq:tensorp01}).

%======================================================
\section{Results}
\begin{figure}[]
%\hspace{-2.5cm}
\centering
\begin{minipage}{14cm}
\includegraphics[origin=c,angle=0,width=7cm,height=6cm]{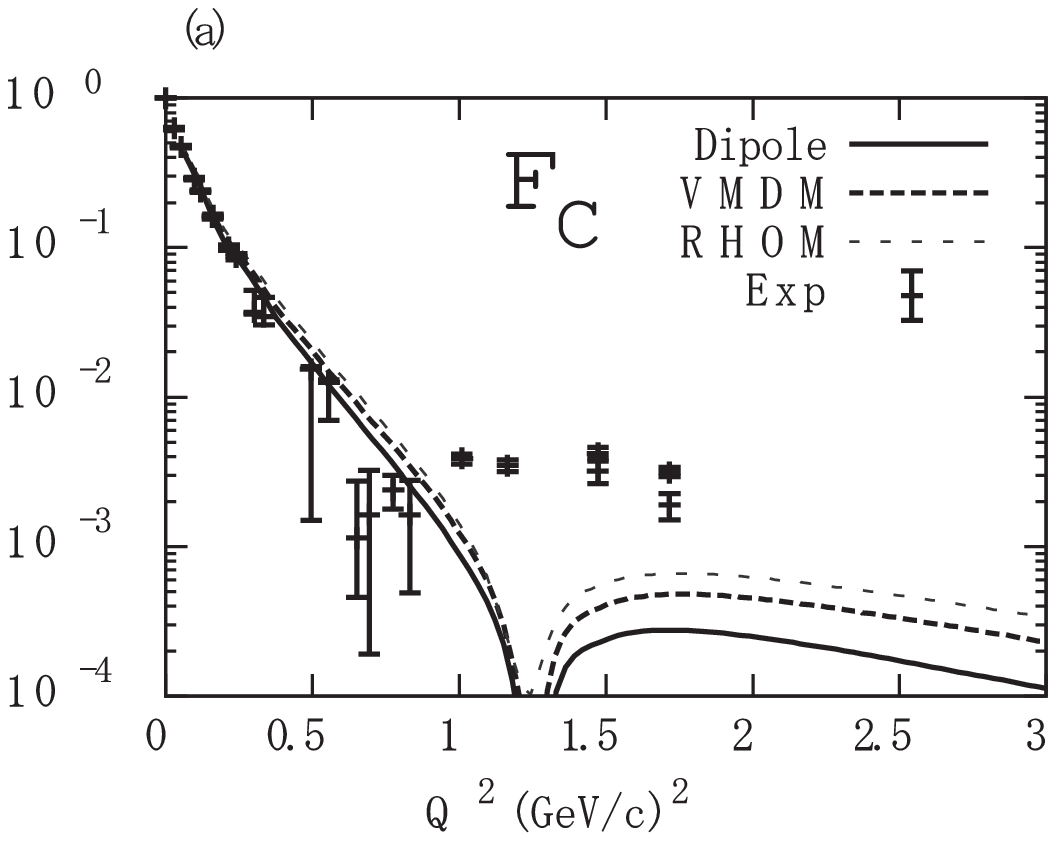}
\includegraphics[origin=c,angle=0,width=7cm,height=6cm]{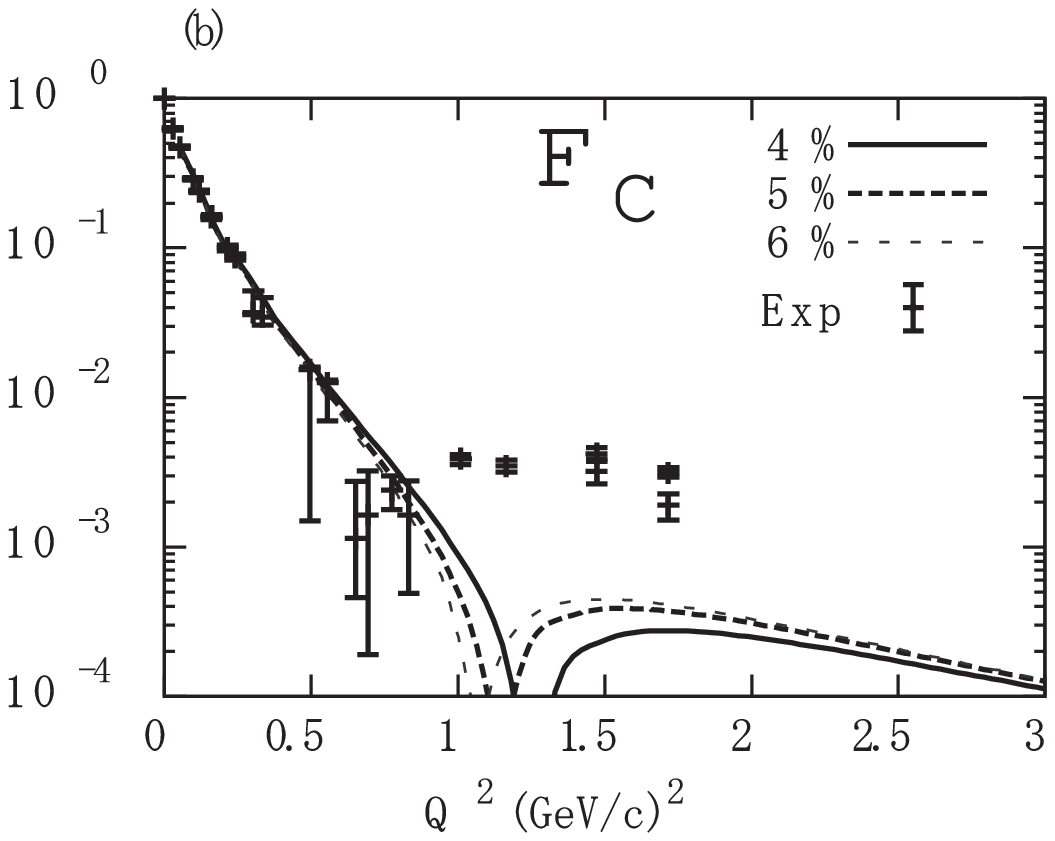}
\caption{The Charge form factor of the deuteron as function of $Q^2$.
The results of the left panel (a) are calculated with the $D$-wave ratio $4\%$
and with three different form factors, Dipole, VMDM and RHOM (see text).
The results of the right panel (b) are calculated with the nucleon form factors Dipole
and with the $D$-wave ratio $4\%$, $5\%$ and $6\%$.
For both cases, the negative energy $P$-wave ratio is fixed to be zero.}
          \label{fig:tpdtd4dvr}
\end{minipage}
\end{figure}

%======================================================
We have calculated the deuteron form factors and tensor polarizations.
Our purpose is to study effects of negative energy components.
Therefore, first we show results of the impulse approximation
without negative energy components.
As shown in Figs.~\ref{fig:tpdtd4dvr}, the results of this impulse 
approximation does not reproduce experimental data.
Then we attempt to improve the agreement by changing nucleon form factors
and $D$-wave probability.
However, as shown in Figs.~\ref{fig:tpdtd4dvr}, the two effects 
do not improve the agreement.

\subsection{Dependence on the intrinsic nucleon form factor \label{dotinff}}
We consider the typical three types of nucleon form factors:
the dipole fit, the vector meson dominance model (VMDM) and
the relativistic harmonic oscillator model (RHOM).
The details of these form factors are presented in \cite{Bondarenko:2002zz}.
The three form factors reproduce the proton charge radius equally well.
However, they differ in the momentum dependence 
and in neutron form factors.
Therefore, our interest is how these differences affect the deuteron form factors.
In order to see the effect of the different form factors,
it is sufficient to show the results which are 
calculated including only positive energy states, 
${}^3S_1^{+}$ and ${}^3D_1^{+}$,
and with the ratio of the $D$-wave being fixed at $4\%$.
(The evaluation of probabilities of the BS amplitudes is discussed in Appendix~\ref{tpotb}.)
For this probability,
we use the same parameter set as used by 
Rupp and Tjon~\cite{rupptjon:1990} for $P_D=4\%$.
In our case, we have additional parameters
$\lambda_{i4}\, (i=1,2,3,4)$ and $\gamma_3$,
which are set equal zero when the $P$-wave is ignored.
 
Results are shown in Fig.~\ref{fig:tpdtd4dvr}-(a) for the charge form factors
of the deuteron,
where solid line is calculated using the dipole form factor,
the dashed line with the VMDM,
and the dotted line with the RHOM,
which are compared with experimental data~\cite{expdat:tensor}.
From this comparison,
we verify that the deuteron form factors are not very sensitive to 
the nucleon form factors.
This result is understandable since the major structure
of the deuteron form factor is determined by the loosely bound nature of the wave function
which makes the form factor fall off very rapidly
as the momentum transfer $Q^2$ is increased.
As compared to the wide spreading structure of the deuteron wave function,
nucleons are regarded as small objects.
Although we have shown the result of the charge form factor only,
other quantities such as magnetic and quadrupole form factors
are also not sensitive to the models of the nucleon form factor.

\subsection{The role of the $D$-wave}
We investigate the role of the $D$-wave in the charge form factor.
The $D$-wave ratio is usually estimated as $4 \sim 6 \%$.
Therefore, we have calculated form factors with
the $D$-wave ratio at $4, 5$ and $6\%$.
We use the parameter sets of  
Rupp and Tjon for
$P_D=4$, $5$ and $6\%$.
As in the previous subsection, parameters
$\lambda_{i4}\, (i=1,2,3,4)$ and $\gamma_3$
are set equal zero.
Results are shown in Fig.~\ref{fig:tpdtd4dvr}-(b).
The solid line is the result obtained with the $D$-wave ratio $4\%$,
the dashed line $5\%$, the dotted line $6\%$.
From this, it is not easy to make a significant improvement
as to reproduce the experimental data with a reasonable range of $D$-wave ratio.   

\subsection{The role of negative energy $P$-waves}
In the previous subsections, it was shown that 
nucleon form factors and the $D$-wave ratio did not improve the 
agreement with the experimental data.
Therefore, in this subsection, 
we investigate the effect of the negative energy 
$P$-waves on
the form factors and tensor polarizations.
We performed calculations by including the two $P$-wave components of
$^1P_1^e$ and $^1P_1^o$ as explained in the section 2.2.
We adopt the nucleon form factor of dipole fit and the $D$-wave ratio $4\%$  

\begin{figure}[]
\centering
\begin{minipage}{14cm}
\hspace{0cm}
\includegraphics[origin=c,angle=0,width=7cm,height=6cm]{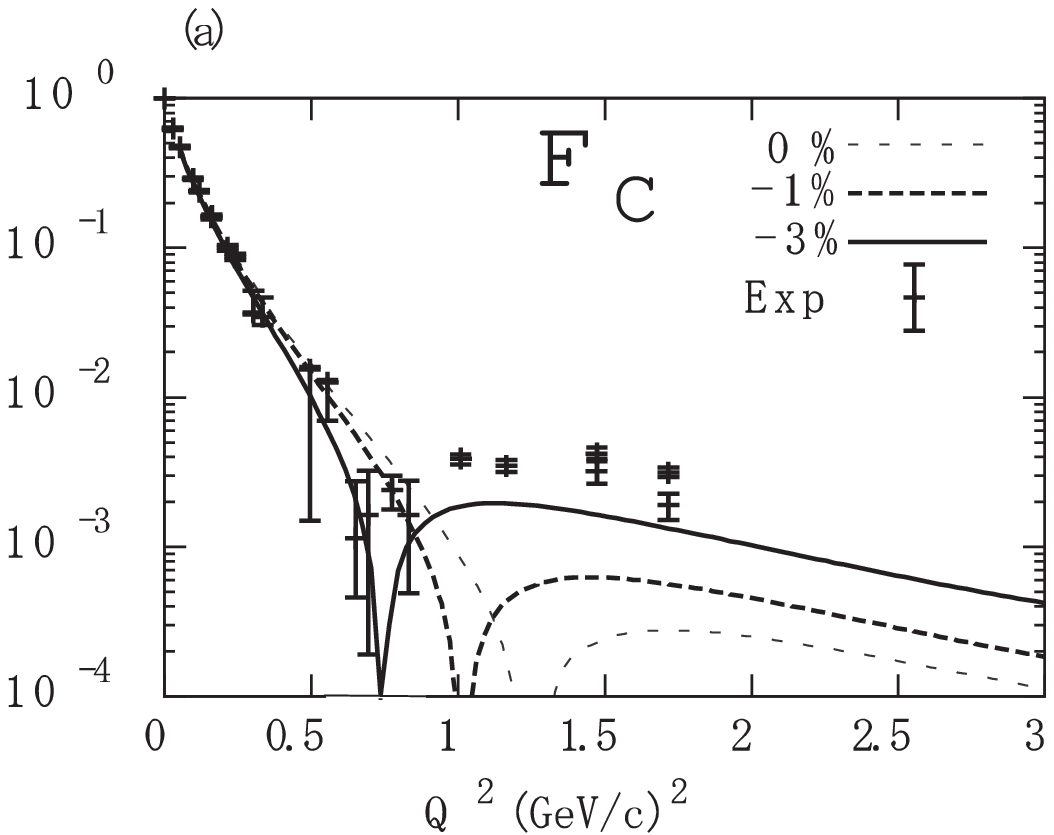}
\hspace{0cm}
\includegraphics[origin=c,angle=0,width=7cm,height=6cm]{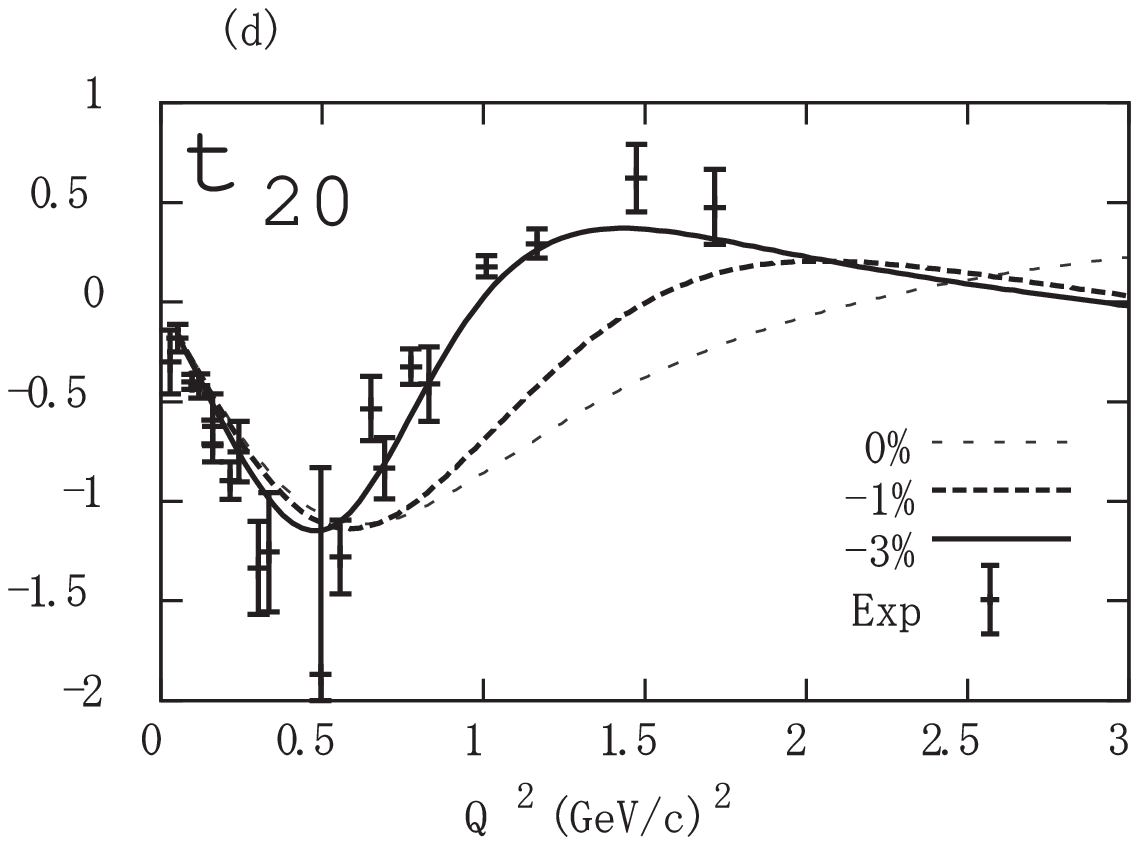}
\end{minipage}
\\
\vspace{0cm}
\centering
\begin{minipage}{14cm} 
\hspace{0cm}
\includegraphics[origin=c,angle=0,width=7cm,height=6cm]{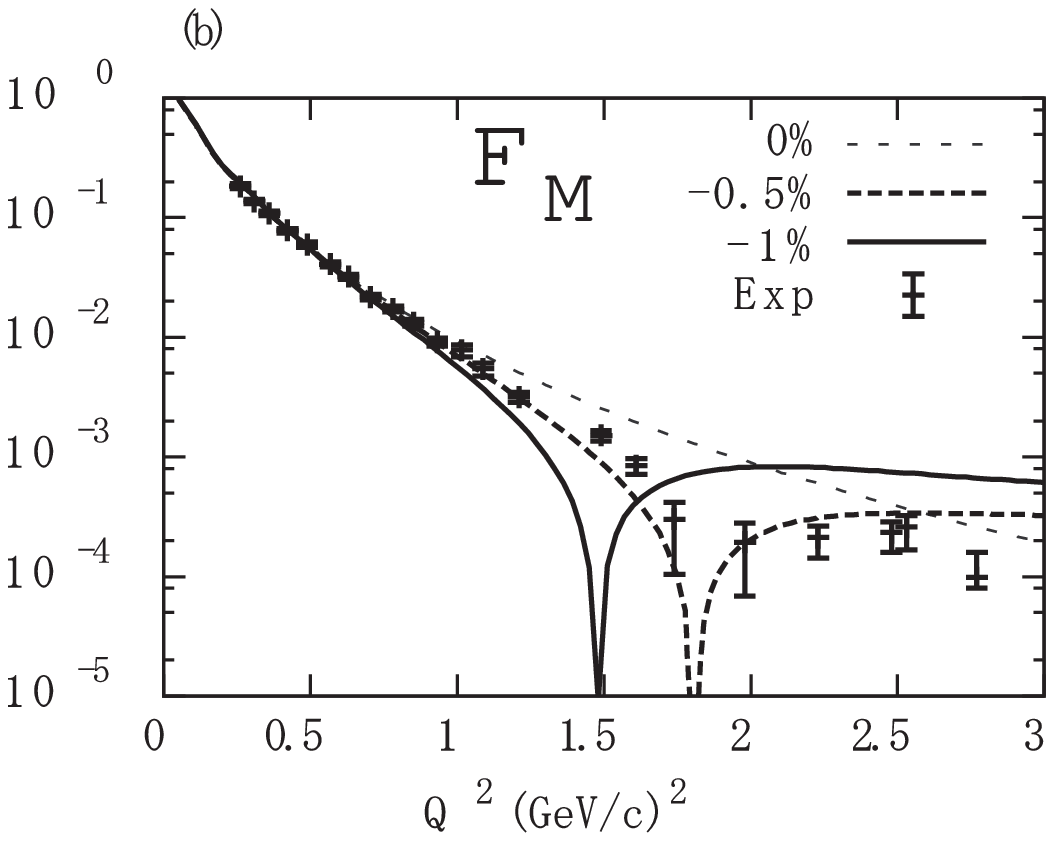}
\hspace{0cm}
\includegraphics[origin=c,angle=0,width=7cm,height=6cm]{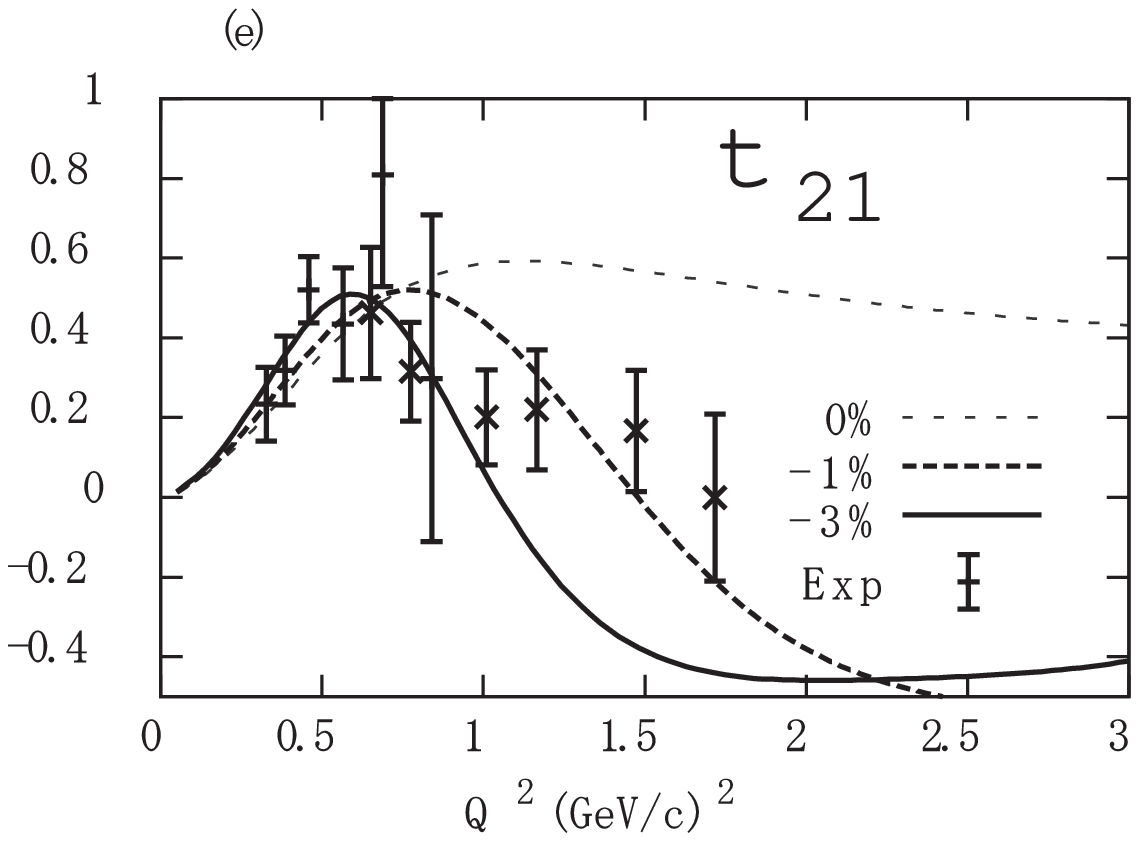}
\end{minipage}
\\ 
\vspace{0cm}
\centering
\begin{minipage}{14cm}
\includegraphics[origin=c,angle=0,width=7cm,height=6cm]{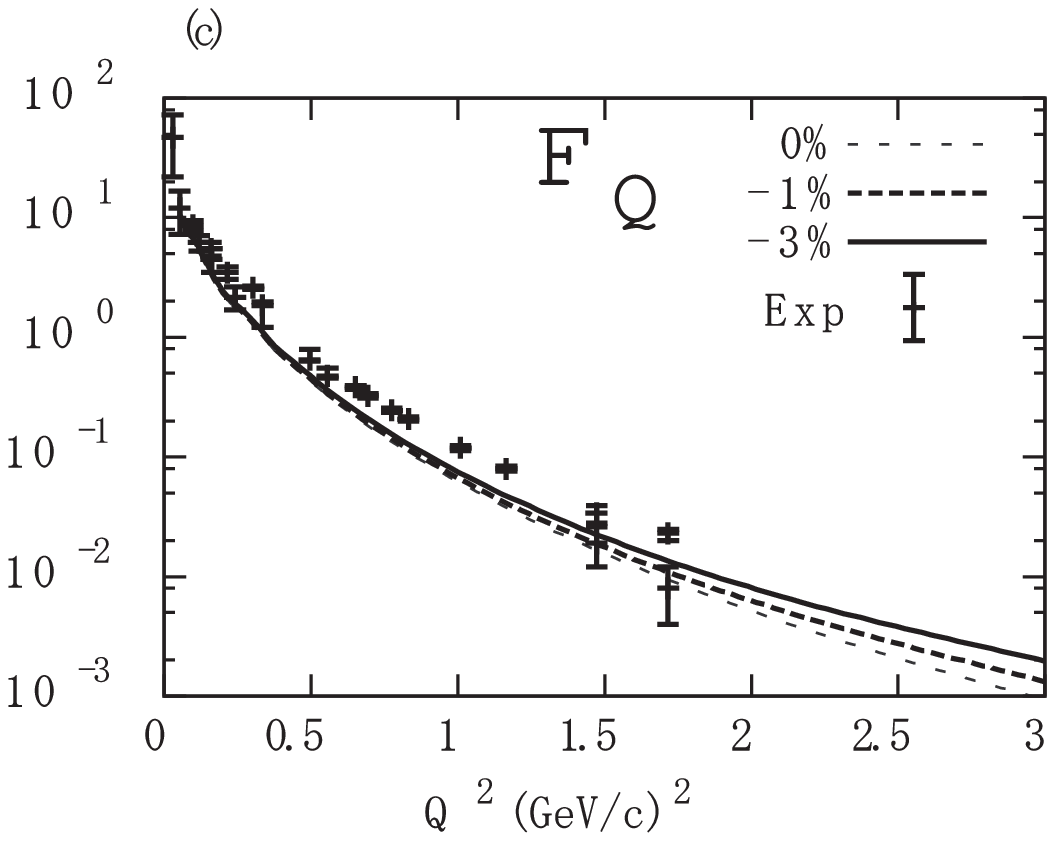}
\includegraphics[origin=c,angle=0,width=7cm,height=6cm]{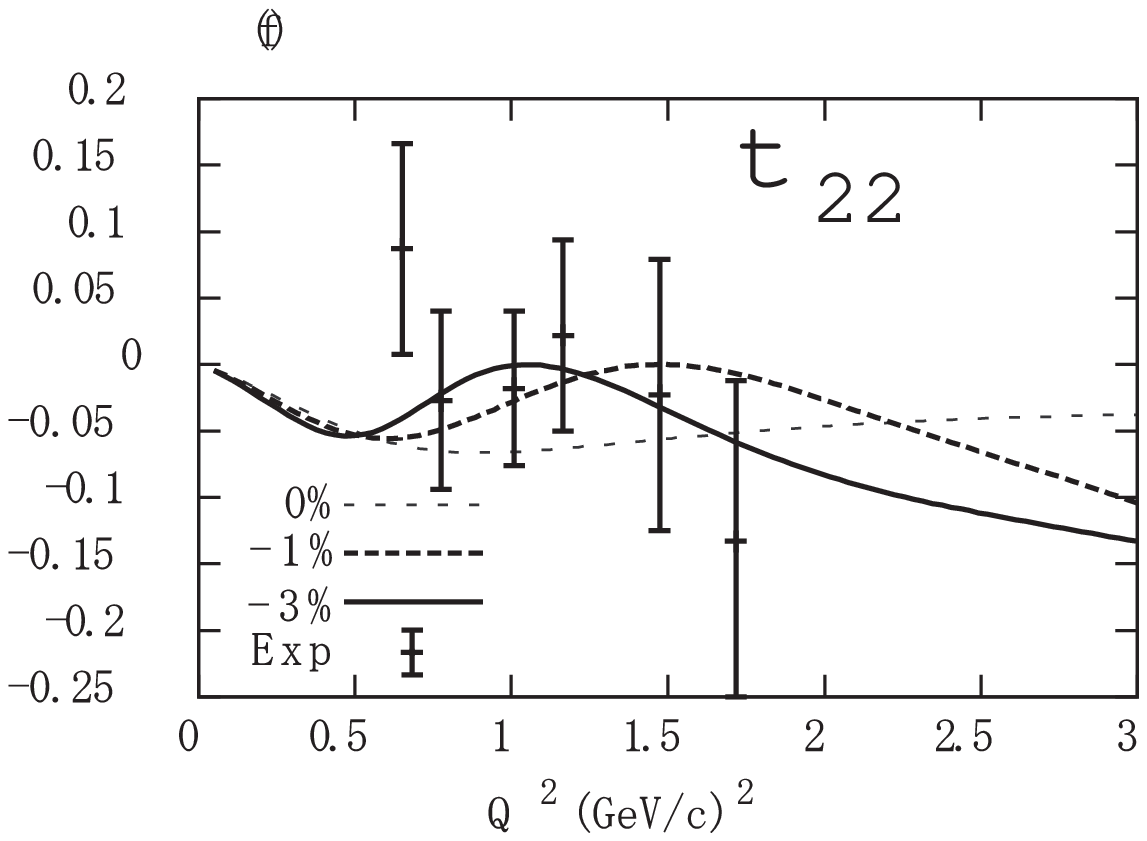}
\caption{Form factors (a)-(c) and tensor polarizations (d)-(f) of the deuteron.
Calculation are performed with the $P$-wave ratio as indicated in each panel.
In all cases the Dipole form factor is used with the 
$D$-wave ratio $4\%$.}
          \label{fig:tpd4p013}
\end{minipage}
\end{figure}

The results are shown in Figs.~\ref{fig:tpd4p013}.
Fig.~\ref{fig:tpd4p013}-(a), (b), (c) show charge, magnetic and quadrupole form factors
respectively,
and Fig.~\ref{fig:tpd4p013}-(d), (e), (f) the $t_{20}$, $t_{21}$ and 
$t_{22}$ tensor polarizations, respectively.

The dotted lines represent the result when the the $P$-wave ratio is $0\%$,
the dashed lines $-1\%$ and the solid lines $-3\%$, except for Fig.~\ref{fig:tpd4p013}-(b). 
For Fig.~\ref{fig:tpd4p013}-(b), the dotted, dashed and solid lines represent the results with  
$P$-wave ratio is $0\%$, $-0.5\%$ and $-1\%$, respectively.
The negative values are called pseudo probabilities 
which are related to the baryon charge~\cite{Zuilhof:1980ae}. 
Details are discussed in Appendix C.
The ratio of negative energy states is controlled by the parameter $u_4$ which  
is chosen
as $u_4 = -3.5, -5.0$ and $-8.5$ for the $P$-wave ratio $-0.5\%$, $-1\%$  
and $-3\%$, respectively.
The corresponding $g$ function parameters are chosen as
$\beta_3 = 0.481$ GeV and $\gamma_3 = -15.0$.
$\gamma_3$ and $\beta_3$ are determined as follows.
We try to reproduce the location of the dip (sign change) of $F_C$, 
firstly by setting $\gamma_3 = -15.0$.
$\beta_3$ is then determined so that $H_{44}(s=M^2)=0$.
%Then we obtain $\beta_3 = 0.481$ GeV.
The rest of the parameters are taken from Rupp and Tjon Ref.~\cite{rupptjon:1990}
of $D$-wave ratio $4\%$.
The data of the charge and quadrupole form factors are taken from Ref.~\cite{expdat:tensor},
those of the magnetic form factor Ref.~\cite{expdat:mff},
and those of the tensor polarizations Refs.~\cite{expdat:tensor,expdat:tensor2}.

We find that better agreement with experimental data is achieved
when we include a finite amount of $P$-wave amplitudes of the negative energy components.
For charge and magnetic form factors, the location of the dip 
and the $Q^2$ dependence especially at high momentum region
$Q^2 > 1$ GeV are significantly improved by including the negative  
energy component.
The amount of the negative energy ratio differs, however.
The best agreement is achieved with $-3 \%$ for the charge form factor,  
while
$-0.5\%$ for the magnetic form factor.
No significant change was found as for the quadrupole form factor.
As for the tensor polarization, the agreement for $t_{20}$ is remarkably
improved when the negative energy ratio is $-3 \%$.
The agreement of the other two components is also
reasonably improved with a finite amount of negative energy components.
The necessity of different amount of negative energy components in order to
reproduce different form factors might be a consequence of
our incomplete treatment of the negative energy components,
where we have ignored negative energy components of
$^3P_1^e$, $^3P_1^o$, $^3S_1^-$ and $^3D_1^-$.
The inclusion of the full set of components
will be a future work. 

\section{Conclusion}
In this paper, we have investigated the effect of negative energy components
on two-nucleon systems.
For that purposes, we have solved the Bethe-Salpeter equation,
including the $^1P_1^e$, $^1P_1^o$ states,
both for scattering states and the bound state.
We have found the parameter sets which reproduce 
electromagnetic properties and the phase shifts of the deuteron. 
On one hand, the inclusion of the negative energy components
improves systematically bound state properties such as the form factors and tensor polarizations
in the relativistic impulse approximation.
On the other hand, using the same parameter set
we have seen that 
the negative energy components affect very little
on scattering properties.
 
Although we have not discussed in detail, we have compared our results
with the previous study based on the non-relativistic formalism.
In the non-relativistic method, the results depend very much on the
approximation, the way how exchange currents are included, and so on.
In the present relativistic method, although our scheme is a
simple impulse approximation, the inclusion of negative energy
components systematically improved the agreement with data.
In this respect, it would be of great interest to investigate
the non-relativistic correspondence of the present relativistic method.

Although the present approach provides formally a systematic method  
to study
two nucleon systems in a relativistic way, actual computations are  
rather
complicated due to the increase in the number of components of the BS
amplitudes and the analytic structure in the complex energy plane.
For the reasons associated with these practical problems, we had to ignore
mathematical rigor
where only part of negative energy components were taken care of.
Nevertheless, our attempt to include for the first time negative energy
components have shown that it indeed improves the agreement
between the theory and experimental data in a systematic manner.
It will be of great interest to further investigate the role of
negative energy components, which is equivalent to
the antiparticle degrees of freedom.
Such dynamics is expected to become important
for more strongly interacting systems such as quark systems
for hadrons.
 
%======================= ===============================
\section*{Acknowledgements}
%======================================================
We thank Valery~V.~Burov and Serge~G.~Bondarenko and Naohide Hamamoto
for fruitful discussions for the relativistic separable approach.
%======================================================

\appendix
\section{Probabilities of the BS amplitude \label{tpotb}}
In this appendix we discuss how probability is defined for 
the BS amplitude.
From the normalization condition Eq.~(\ref{normalization2}) when $\mu=0$,
we can obtain the normalization condition for the radial part as follows,
%begin{eqnarray}
%=(\sqrt{s},0)=(M,0)\,,
%end{eqnarray}
\begin{eqnarray}
&&\int \, d k_0\,\int\,d|\bk|  |\bk|^2
\Gamma^{\dagger}(k_0,\nrmk)
\Bigl[
\frac{\partial}{\partial {\sqrt{s}}} 
{S}(k_0,\nrmk;s)
\Bigr]_{\sqrt{s}=M}
 \Gamma(k_0,\nrmk) =
-2M \,,
\label{tpotb1}
\end{eqnarray}
where the integration over angles has been done
and $\Gamma(k_0,\nrmk)$ is the radial part of the vertex function.
%and in the case of $J=1$ deuteron case
For the deuteron of $J=1$,
$S$ is the two nucleon propagator define by Eqs.~(\ref{2nprgtr}) and (\ref{eq:tnps})
which is given as a matrix form
for eight channels. 

Substituting the definition of the radial part of the BS amplitude 
\begin{eqnarray}
\Gamma(k_0,\nrmk)=S^{-1}\phi(k_0,\nrmk)
\end{eqnarray}
obtained from Eq.~(\ref{eq:BSA-vertex})
into Eq.~(\ref{tpotb1}), we find
\begin{eqnarray}
&&\int\, d k_0\,\int\,d|\bk|  |\bk|^2
\phi^{\dagger}(k_0,\nrmk)
\bigl[
%\
S^{-1}
\bigr]_{\sqrt{s}=M}
\Bigl[
\frac{\partial}{\partial{\sqrt{s}}}S
\Bigr]_{\sqrt{s}=M}
\bigl[
S
\bigr]_{\sqrt{s}=M}
\phi(k_0,\nrmk)
\nn\\
&=&\sum \limits_{\alpha}\int\, d k_0\,\int\,d|\bk|  |\bk|^2
\phi_{\alpha}^{\ast}(k_0,\nrmk)
\omega_{\alpha\beta}
\phi_{\beta}(k_0,\nrmk)
\nn\\
&=&-2M\,.
\end{eqnarray}
In the second line we have used eight components explicitly,
\begin{eqnarray}
\omega_{\alpha\beta}&=&
\left[
\bigl[
%\
S^{-1}
\bigr]_{\sqrt{s}=M}
\Bigl[
\frac{\partial}{\partial{\sqrt{s}}}S
\Bigr]_{\sqrt{s}=M}
\bigl[
S
\bigr]_{\sqrt{s}=M}
\right]_{\alpha\beta}
\nonumber
\\
&=&
\left(
    \begin{array}{@{\,}cccccccc@{\,}}
    -\frac{M}{2}+E_k & 0 & 0 & 0 & 0 & 0 & 0 & 0\\
    0      &  -\frac{M}{2}+E_k     & 0 & 0 & 0 & 0 & 0 & 0\\
    0 & 0 &   -\frac{M}{2}-E_k & 0 & 0 & 0 & 0 & 0\\
    0 & 0 & 0 &  -\frac{M}{2}-E_k & 0 & 0 & 0 & 0\\
    0 & 0 & 0 & 0 & -\frac{M}{2} & 0 &0  &0 \\
    0 & 0 & 0 & 0 & 0 & -\frac{M}{2} &0  &0 \\
    0 & 0 & 0 & 0 & 0 & 0 & -\frac{M}{2} &0 \\
    0 & 0 & 0 & 0 & 0 & 0 &0  &-\frac{M}{2}  
    \end{array}
\right)
%\]
\nn\,,
\end{eqnarray}
\begin{eqnarray}
%\[
\phi(k_0,\nrmk)= 
 \left(
    \begin{array}{@{\,}c@{\,}}
   \phi_{{}^3S_1^{+}}(k_0,\nrmk) \\
   \phi_{{}^3D_1^{+}}(k_0,\nrmk)  \\
   \phi_{{}^3S_1^{-}}(k_0,\nrmk)  \\
   \phi_{{}^3D_1^{-}}(k_0,\nrmk)  \\
   \phi_{{}^1P_1^{e}}(k_0,\nrmk)  \\
   \phi_{{}^3P_1^{o}}(k_0,\nrmk) \\
   \phi_{{}^1P_1^{o}}(k_0,\nrmk) \\
   \phi_{{}^3P_1^{e}}(k_0,\nrmk) 
  \end{array}
\right)
%\]
\nn\,,
\end{eqnarray}
\begin{eqnarray}
\phi^{\dagger}(k_0,\nrmk)
 =
 \bigl(
   \phi^{\ast}_{{}^3S_1^{+}}(k_0,\nrmk),
   \phi^{\ast}_{{}^3D_1^{+}}(k_0,\nrmk),
   \phi^{\ast}_{{}^3S_1^{-}}(k_0,\nrmk),
   \phi^{\ast}_{{}^3D_1^{-}}(k_0,\nrmk),
\nonumber
\\
   \phi^{\ast}_{{}^1P_1^{e}}(k_0,\nrmk),
   \phi^{\ast}_{{}^3P_1^{o}}(k_0,\nrmk),
   \phi^{\ast}_{{}^1P_1^{o}}(k_0,\nrmk),
   \phi^{\ast}_{{}^3P_1^{e}}(k_0,\nrmk)
\bigr)\,.
 \nonumber
\end{eqnarray}
From these equations,
we can define the probability of finding the $\alpha$-state in the total deuteron state 
by the following expression,
\begin{eqnarray}
P_{\alpha}=\frac{1}{N}\int\, d k_0\,\int\,d|\bk|  |\bk|^2
 \omega_{\alpha}|\phi_{\alpha}(k_0,\nrmk)|^2\,,
\label{probability8}
\end{eqnarray}
where $N$ is determined by 
\begin{eqnarray}
\sum \limits_{\alpha} P_{\alpha}=1\,\,
%(\alpha=1\thicksim 8)\,.
(\alpha=^3S_1^{+}, ^3D_1^{+}, ^3S_1^{-}, ^3D_1^{-}, ^1P_1^{e},
 ^1P_1^{o}, ^3P_1^{e}, ^3P_1^{o})\,. 
\end{eqnarray}
$P_{\alpha}$ is often called as {\it``}pseudo probability'' which is normalized
by the baryon charge \cite{Zuilhof:1980ae}.
When $\alpha=^3S_1^{-}, ^3D_1^{-}, ^1P_1^{e},
 ^1P_1^{o}, ^3P_1^{e}, ^3P_1^{o}$,
the sign of $P_{\alpha}$ is minus in accordance with the sign of 
$\omega_{\alpha\alpha} (\alpha=1, \cdots, 8).$

For example in the case of $^3S_1^{+}$
\begin{eqnarray}
P_{^3S_1^{+}}&=&\frac{1}{N}\int\, d k_0\,\int\,d|\bk|  |\bk|^2
 \omega_{\alpha}|\phi_{\alpha}(k_0,\nrmk)|^2
\nonumber
\\
&=&\frac{1}{N}\int\, d k_0\,\int\,d|\bk|  |\bk|^2
(-\frac{M}{2}+E_k)\frac{1}{\left((M/2-E_k+i\epsilon)^2-k_0^2\right)^2}\
\Gamma_{^3S_1^{+}
}(k_0,\nrmk)^2\,.
\nonumber
\end{eqnarray}
 
If we assume that $\Gamma_{^3S_1^{+}}(k_0,\nrmk)$ has no pole in the $k_0$-plane
(when there is a pole, we need careful treatment of the location
of the pole of $\Gamma$.),
we have the only one pole from the propagator $S_+$, namely at $k_0=\frac{M}{2}-E_k+i\epsilon$.  
As a result we can obtain
\begin{eqnarray}
P_{^3S_1^{+}}&=&\frac{1}{N'}\int\,d|\bk|  |\bk|^2
\left(
\frac{\Gamma_{^3S_1^{+}
}(\overline{k_0},\nrmk)}{(M/2-E_k)}\,
\right)^2\,,
\label{relasw}
\end{eqnarray}
where $\overline{k_0}=M/2-E_k$.
Here we have used
\begin{eqnarray}
\int\, d k_0\,\left(
\frac{1}{k_0^2-\overline{k_0}^2}
\right)^2
{\Gamma_{^3S_1^{+}}(k_0,\nrmk)}^2
&=&
\int\, d k_0\,
\frac{\partial}{\partial{\overline{k_0}^2}}
\left(
\frac{1}{k_0^2-\overline{k_0}^2}
\right)
{\Gamma_{^3S_1^{+}}(k_0,\nrmk)}^2
\nonumber
\\
&=&
\frac{\partial}{\partial{\overline{k_0}^2}}
\left(
2\pi\,i
\frac{1}{2\overline{k_0}}
{\Gamma_{^3S_1^{+}}(\overline{k_0},\nrmk)}^2
\right)
\nonumber
\\
&=&
%\left(
-\frac{2\pi\,i}{4\overline{k_0}^3}
{\Gamma_{^3S_1^{+}}(\overline{k_0},\nrmk)}^2\,.
%\right)
\end{eqnarray}

From Eq.~(\ref{relasw}) we can consider that
\begin{eqnarray}
%\left(
\frac{\Gamma_{^3S_1^{+}
}(\overline{k_0},\nrmk)}{(M/2-E_k)}\,
%\right)
\label{relationnonrelarela}
\end{eqnarray}
corresponds to the momentum space wave function in the non-relativistic framework.
From the Schrodinger equation
\begin{eqnarray}
(H_0+V)\psi=E\psi\,,
\end{eqnarray}
the wave function can be expressed as  
\begin{eqnarray}
\psi=\frac{1}{E-H_0}V\psi\equiv\frac{1}{E-H_0}\Gamma\,.
\label{nonrelasolution}
\end{eqnarray}
Here we have used $V\psi=\Gamma$, which is obtained from Eq.~(\ref{eq:BSA-vertex}) and 
Eq.~(\ref{eq:BSA-eq}). 
Comparing Eq.~(\ref{relationnonrelarela}) and Eq.~(\ref{nonrelasolution}),
we can immediately understand that Eq.~(\ref{relationnonrelarela}) corresponds 
to the Schrodinger wave function.
\section{Negative energy contributions in a static potential\label{sapole}}
In this appendix, we consider the BS equation in a static separable potential,
where we study the negative energry contribution for bound and scattering states.

Let us start with the discussion of the scattering states.
To be definite, we consider the scattering of the $^3S_1^{+}$ channel,
where intermediate states are restricted to $^3S_1^{+}$, $^3D_1^{+}$, $^1P_1^{e}$ and $^1P_1^{o}$,
as used in this work.
Hence in the BS equation (\ref{eq:sep01}),
the in and out states are $\alpha=\beta={^3S_1^{+}}$.
The other channels enters intermediate states labeled by $\gamma$ and $\delta$.
The propagators for the negative energy components are $S_e$ and $S_o$
which are the combination of $S_{+-}$ and $S_{-+}$
as given by the Eqs.~(\ref{eq:tnps}).
The location of the poles of $S_e$ and $S_o$ are given in the Fig.~\ref{fig:cps}
in the $k_0$-plane. 

In the static approximation for the $g$-function,
we set $k_{0}=0$:
\begin{eqnarray}
&&g_{4}^{^1P_1^e}(k_0,|{\bf k}|)=\frac{|{\bf k}|}
{(k_0^2-{\bf k}^2-\beta_{3}^2+i\epsilon)^{2}} \stackrel{{k_0}=0}{\longrightarrow} \frac{|{\bf k}|}
{({\bf k}^2+\beta_{3}^2)^{2}},\\\nonumber
&&g_{4}^{^1P_1^o}(k_0,|{\bf k}|)=\gamma_{3}\frac{ p_0}{m}\frac{|{\bf k}|}
{(k_0^2-{\bf k}^2-\beta_{3}^2+i\epsilon)^{2}}\stackrel{{k_0}=0}{\longrightarrow} 
0\,.
\end{eqnarray}
Substituting these equations, the $k_0$ integral for the negative energy contributions
becomes 
\begin{eqnarray}
\int\,d k_0\,
\frac{|\bf k|^2}
{({\bf k}^2+\beta_{3}^2)^{4}}
\{S_{+-}(k_0,|\mbf{k}|)+S_{-+}(k_0,|\mbf{k}|)\}=0\,.
\end{eqnarray}
Here, we used the facts that
$g$-functions are indipendent of $k_0$
and the poles of $S_{+-}$ and $S_{-+}$ are located in the same side
with respect to the real axis on the complex plane.
In this way, we can show that the contribution of intermediate states of 
negative energy do not contribute to the scattering amplitude.

Next, let us consider the bound states.
For this purpose we consider the BS equation for the bound state amplitude
of each channel ($^3S_1^{+}$, $^3D_1^{+}$, $^1P_1^{e}$, $^1P_1^{o}$).
In particular,
we need to look at the one of negative energy, say, $\alpha={{{}^1}P_1^e}$
and investigate whether it survives or not.
Then, the BS equation we should solve is 
\begin{eqnarray}
\phi_{{^1P_1^e}}(p_0,|\bp|) &=&
\int\,d k_0\,\mbf{k}^2\,d|\mbf{k}|\,
\sum\limits_{\beta\gamma}\,
S_{{^1P_1^e}\beta}(p_0,|\mbf{p}|)\,
V_{\beta\gamma}(p_0,|\bp|,k_0,|\mbf{k}|)\,
\phi_{\gamma}(k_0,|\mbf{k}|)
\,
\nonumber
\\
&=&
\sum\limits_{\beta\gamma\delta}\,
\sum_{i,j,k,l=1}^{N}\,
c_l\lambda_{ij}\lambda_{kl}
S_{{^1P_1^e}\beta}(p_0,|\mbf{p}|)\,
g_{i}^{(\beta)}(p_0,|\mbf{p}|)
\nonumber
\\
&\times
&
\int\,d k_0\,\mbf{k}^2\,d|\mbf{k}|\,
S_{\gamma\delta}(k_0,|\mbf{k}|)
g_{j}^{(\gamma)}(k_0,|\mbf{k}|)
g_{k}^{(\delta)}(k_0,|\mbf{k}|)\,.
\end{eqnarray}
Here, unlike the BS equation of the form $\psi=\phi+SV\psi$ for scattering states,
there is no inhomogeneous term correponding to the incoming plane wave $\phi$.
In the second line, we use the separable ansatz soulution of Eq.~(\ref{eq:sep06}).
Using the static approximation for the $g$-function,
the $k_0$ integral for the negative energy components
becomes zero as in the case of the scattering states.
However, the part related to the positive energy components remains.
Therfore, the BS amplitude $\phi_{{^1P_1^e}}(p_0,|\bp|)$ does not become zero
in the static approximation.
For $\alpha={{{}^1}P_1^o}$ case, we can discuss in the same way.
Therefore, we should expand the BS amplitude by the partial wave 
including the negative energy components.

From these discussions for scattering and bound states in the static approximation,
we can justify the prescription of the pole locations as discussed in section (\ref{section:cwnec}).
\end{document}